\documentclass[english,notitlepage,nofootinbib]{revtex4-1}
\usepackage[T1]{fontenc}
\usepackage[latin9]{inputenc}
\usepackage{color}
\usepackage{verbatim}
\usepackage{amsmath}
\usepackage{placeins}
\usepackage{stackrel}
\usepackage{graphicx}
\usepackage{esint}
\usepackage{babel}
\usepackage{url}
\usepackage{hyperref}
\usepackage{subfig} 
\usepackage{float}
 \usepackage[subfigure,titles]{tocloft}
\usepackage{geometry}
\usepackage{makecell}
\usepackage{multirow}

\makeatletter
\renewcommand\@makecaption[2]{%
  \par
  \vskip\abovecaptionskip
  \begingroup
   \small\rmfamily
    \begingroup
     \samepage
     \flushing
     \let\footnote\@footnotemark@gobble
     \@make@capt@title{#1}{#2}\par
    \endgroup
  \endgroup
  \vskip\belowcaptionskip
}
\makeatother

\def\gs{\mathrel{
   \rlap{\raise 0.511ex \hbox{$>$}}{\lower 0.511ex \hbox{$\sim$}}}}
\def\ls{\mathrel{
   \rlap{\raise 0.511ex \hbox{$<$}}{\lower 0.511ex \hbox{$\sim$}}}}

\begin{document}

\title{Naturalness in testable type II seesaw scenarios}

\date{\today}

\author{P. S. Bhupal Dev$^a$, Clara Miralles Vila$^b$, Werner Rodejohann$^b$}

\affiliation{
$^a$Department of Physics and McDonnell Center for the Space Sciences, Washington University, 
St.\ Louis, MO 63130, USA\\
$^b$Max-Planck-Institut f\"ur Kernphysik, Saupfercheckweg 1, D-69117 Heidelberg,
Germany}
\begin{abstract}
\noindent
New physics coupling to the Higgs sector of the Standard Model can lead to dangerously large 
corrections to the Higgs mass. We investigate this problem in the type II seesaw model for neutrino mass, 
where a weak scalar triplet is introduced. 
The interplay of direct and indirect constraints on the type II seesaw model with its 
contribution to the Higgs mass is analyzed. The focus lies on testable triplet masses and 
(sub) eV-scale triplet vacuum expectation values. We   
identify scenarios that are testable in collider and/or lepton flavor violation experiments, while satisfying the Higgs naturalness criterion.

\end{abstract}
\maketitle

\section{\label{sec:intro}Introduction}
The absence of any compelling new physics signal at high energy and high sensitivity experiments strengthens the case for studying the 
issue of the hierarchy problem and naturalness. New physics coupling to the Higgs sector of the Standard Model (SM) is expected to give radiative corrections to the 
Higgs mass, and according to the naturalness condition, these corrections must not exceed the 
measured value of the Higgs mass. Various such possibilities exist for new physics coupling to the SM Higgs boson. 
In search for a motivated ansatz, one notes 
that neutrino physics is an attractive starting point, since it remains 
the only new physics beyond the SM that has been observed in laboratory experiments.  
Indeed, most mechanisms that generate neutrino mass include couplings of new particles with the SM Higgs doublet. 
In those scenarios, naturalness considerations can provide useful constraints on the parameters responsible for 
neutrino mass, see e.g.~Refs.~\cite{Vissani:1997ys,Casas:2004gh,Abada:2007ux,Xing:2009in,Farina:2013mla,Clarke:2015gwa,
Fabbrichesi:2015zna,Clarke:2015hta,Chabab:2015nel,Haba:2016zbu,Clarke:2016jzm,
Haba:2016lxc, Bambhaniya:2016rbb,Rink:2016knw}.  

This paper will focus on the minimal type II seesaw mechanism~\cite{Schechter:1980gr,Cheng:1980qt,Lazarides:1980nt,Mohapatra:1980yp}, in which a weak scalar triplet is 
responsible for neutrino mass. Unlike the type I or III seesaw mechanisms, where the radiative contribution to the 
Higgs mass is suppressed by the lightness of neutrino mass, and thus very large masses (up to $10^7$ GeV) of the 
new particles (fermion singlets and triplets) are allowed, the type II seesaw can pose problems 
already when the triplet mass is of order TeV. These values are testable at colliders~\cite{Perez:2008ha, Melfo:2011nx, kang:2014jia, Chen:2014qda, Han:2015hba, Dev:2016dja, Mitra:2016wpr} and with lepton 
flavor violation (LFV)~\cite{Akeroyd, Fukuyama, Chakrabortty, Chakrabortty:2015zpm}. Apart from the direct LHC and LFV constraints, the full type II seesaw model faces several indirect constraints such as unitarity, 
vacuum stability and perturbativity~\cite{Dey:2008jm, Gogoladze:2008gf, Arhrib, Chun, Chao:2012mx, Dev:2013ff, Kobakhidze:2013pya, Chakraborty:2014xqa, Bonilla, Chabab:2015nel, Haba:2016zbu, Das:2016bir, Xu:2016klg, Biswas:2017tnw}, which should be satisfied up to the Planck scale, if we do not assume any other new physics. 
In this work, we investigate the interplay of all those direct and indirect constraints in light of the naturalness  criterion. We focus on the most appealing scenario, namely testable values of the triplet mass around TeV and (sub) eV values of the triplet vacuum expectation value (VEV). 

The rest of the paper is built up as follows: in Section \ref{sec:formalism} we review the formalism of the minimal type II seesaw 
model and its connection to the Higgs sector of the SM, including the correction to the Higgs mass. 
Section \ref{sec:Constraints} summarizes the various direct and indirect constraints on the model. 
In Section \ref{sec:res} the details and results of our numerical analysis are presented, including 
implications for LFV and Higgs decays $h \to \gamma\gamma$ and $h \to Z\gamma$. 
We conclude in Section \ref{sec:conc}.

\section{\label{sec:formalism}The Minimal Type II SeeSaw Model}

\subsection{\label{sec:Higgses}Basics}

In the minimal type II seesaw model, the SM particle content is enlarged by the addition of a weak triplet complex scalar field $\Delta$, which transforms as $(\textbf{3},1)$ under the $SU(2)_L \times U(1)_Y$ electroweak (EW) gauge group:
\begin{equation}
\Delta \ = \ 
\begin{pmatrix}
   \delta^+ / \sqrt{2} & \delta^{++}\\
   \delta^0 & -\delta^+ / \sqrt{2}
  \end{pmatrix} \, .
\end{equation}
So the scalar sector of the model consists of $\Delta({\bf 3},1)$ and the $SU(2)_L$-doublet Higgs field $\Phi$ which transforms as ({\bf 2},1/2): 
\begin{equation}
\Phi \ = \ 
\begin{pmatrix}
\phi^+ \\
\phi^0
\end{pmatrix} \, .
\end{equation}
The minimal Lagrangian for this model is given by
\begin{equation}
	\mathcal{L}\ = \ \mathcal{L}_\mathrm{kinetic}+\mathcal{L}_Y-\mathcal{V}(\Phi,\Delta)\,,
\label{eq:Lag}
\end{equation}
where the kinetic and Yukawa interaction terms are, respectively, 
\begin{align}
\mathcal{L}_\mathrm{kinetic} & \ = \ \mathcal{L}_\mathrm{kinetic}^\mathrm{SM}+\mathrm{Tr}\left[ (D_{\mu} \Delta)^{\dagger} (D^{\mu} \Delta) \right] \, , \label{eq:lag1} \\
\mathcal{L}_Y & \ = \ \mathcal{L}_Y^\mathrm{SM} - \left(Y_{\Delta}\right)_{ij}L_{Li}^TCi\sigma_2\Delta L_{Lj} + \mathrm{h.c.} \label{eq:lag2}
\end{align}
Here $\sigma_2$ is the second Pauli matrix, $C$ is the Dirac charge conjugation matrix with respect to the Lorentz group and $D_{\mu}\Delta$ is the covariant derivative of the scalar triplet, given by
  \begin{equation}
  D_{\mu}\Delta \ = \ \partial_{\mu}\Delta+i\frac{g}{2}[\sigma^a W^a_{\mu},\Delta] + \frac{g'}{2}B_{\mu}\Delta \qquad \qquad (a=1,2,3), 
  \end{equation}
  where $g$ and $g'$ are the weak and hypercharge interaction couplings, respectively. 
  
The scalar potential in Eq.~\eqref{eq:Lag} can be written as~\cite{Dev:2013ff}
  \begin{align}
  \mathcal{V} (\Phi, \Delta) \ = \ & -\mu^2 \Phi^{\dagger} \Phi + \frac{\lambda}{2} (\Phi^{\dagger} \Phi)^2 + M_{\Delta}^2 \mathrm{Tr}(\Delta^{\dagger}\Delta) + \frac{\lambda_1}{2}\left[\mathrm{Tr}(\Delta^{\dagger}\Delta) \right]^2 \nonumber\\
  & + \frac{\lambda_2}{2}\left( \left[\mathrm{Tr}(\Delta^{\dagger}\Delta)\right]^2 - \mathrm{Tr}\left[(\Delta^{\dagger}\Delta)^2\right]\right) 
  + \lambda_4 (\Phi^{\dagger} \Phi)\mathrm{Tr}(\Delta^{\dagger}\Delta) + \lambda_5 \Phi^{\dagger} [\Delta^{\dagger},\Delta]\Phi \nonumber\\ 
  & + \left( \frac{\Lambda_6}{\sqrt{2}} \Phi^T i \sigma_2 \Delta^{\dagger} \Phi + \mathrm{h.c.}\right) \, .
  \label{Eq:ScalarPotentialSWII}
  \end{align}
The coupling $\Lambda_6$ is a dimension-full parameter, with mass dimension one. The coupling constants $\lambda_i\ (i=1,2,4,5)$ can be chosen to be real through a phase redefinition of the field $\Delta$. The parameters $\mu^2$ and $\lambda$ are chosen to be positive to ensure the spontaneous EW symmetry breaking 
of the $SU(2)_L \times U(1)_Y$ gauge group to $U(1)_{\rm em}$ through the Higgs mechanism, when the neutral component of $\Phi$ acquires a VEV, $\langle \phi^0\rangle=v/\sqrt 2$. This induces a tadpole term for the $\Delta$ field via the $\Lambda_6$ term in Eq.~\eqref{Eq:ScalarPotentialSWII}, thereby generating a non-zero VEV for its neutral component, $\langle \delta^0 \rangle = v_{\Delta}/ \sqrt{2}$.  
  
Minimizing the scalar potential\footnote{See Ref.~\cite{Xu:2016klg} for a recent analysis on 
the condition that the assumed minimum is the global one.} 
with respect to $\Phi$ and $\Delta$ gives 
\begin{align}
m_{\Phi}^2 & \ = \ \frac{1}{2} \lambda v^2 - \Lambda_6 v_{\Delta} + \frac{1}{2}(\lambda_4-\lambda_5)v_{\Delta}^2\,,\\
M_{\Delta}^2 & \ = \ \frac{1}{2} \frac{\Lambda_6v^2}{v_{\Delta}} -\frac{1}{2}(\lambda_4-\lambda_5)v^2-\frac{1}{2}\lambda_1 v_{\Delta}^2\, . 
\label{Eq:EWSB2}
\end{align}
Note that the triplet VEV contributes to the $W$ and $Z$ boson masses, and hence, to the EW $\rho$-parameter at tree-level through the kinetic term~\eqref{eq:lag1}. On the other hand, the EW precision data does not allow the $\rho$-parameter to deviate much from the SM value of 1; from a recent global fit, $\rho=1.00037\pm 0.00023$~\cite{Olive:2016xmw}. This implies $v_\Delta/v<0.02$ or $v_{\Delta} \lesssim 5~\mathrm{GeV}$. In this limit of $v_\Delta\ll v$, we obtain from Eq.~\eqref{Eq:EWSB2}: 
\begin{equation}\label{eq:vD}
v_{\Delta} \ = \ \frac{\Lambda_6 v^2}{2 M_{\Delta}^2 + v^2(\lambda_4 - \lambda_5)}\,.
\end{equation}

From the Yukawa Lagrangian~\eqref{eq:lag2}, we find that the triplet VEV gives rise to a  Majorana mass term for the neutrinos: 
\begin{equation}
\mathcal{L}_Y \ \supset \ - \left(Y_{\Delta}\right)_{ij}L_i^TCi\sigma_2\Delta L_j + \mathrm{h.c.} \quad \longrightarrow \quad -\frac{v_{\Delta}}{\sqrt{2}} \left(Y_{\Delta}\right)_{ij}\nu_{Li}^TC \nu_{Lj} + \mathrm{h.c.}
\end{equation}
The resulting Majorana neutrino mass matrix is given by 
\begin{equation}
(M_{\nu})_{ij} \ = \ \sqrt{2} v_{\Delta} (Y_{\Delta})_{ij}\,,
\label{Eq:MassMatrix}
\end{equation}
with $v_\Delta$ given by Eq.~\eqref{eq:vD}. In the limit $M_{\Delta}^2 \gg v^2$ or $\lambda_4\simeq \lambda_5$, Eq.~(\ref{Eq:MassMatrix}) becomes
\begin{equation}
M_{\nu} \ \simeq \ \frac{\lambda_6 v^2}{\sqrt{2} M_{\Delta}} Y_{\Delta}\,,
\end{equation}
where we have defined the dimensionless parameter $\lambda_6 \equiv \Lambda_6 / M_{\Delta}$. In the diagonal charged lepton basis, $M_\nu$ can be diagonalized as 
\begin{equation}
M_{\nu} \ = \  U^* \mathrm{diag}(m_1,m_2,m_3) U^{\dagger}\,,
\end{equation}
where $U$ is the Pontecorvo-Maki-Nakagawa-Sakata (PMNS) mixing matrix, parameterized by three mixing angles $\theta_{12,23,13}$, one Dirac phase $\delta$ and two Majorana phases $\alpha_{1,2}$~\cite{Olive:2016xmw}. Thus, the structure of the Yukawa coupling matrix $Y_{\Delta}$ is 
constrained by low-energy neutrino oscillation data. With $v_\Delta$ of eV or below, current neutrino mass 
constraints imply Yukawa couplings of order one. The measured 
small neutrino mass-squared differences and large lepton mixing angles further imply that there is no strong hierarchy 
in $Y_\Delta$~\cite{Dev:2013ff}, and hence, values of order one for all entries are an appealing and consistent assumption. 
In this case, and further assuming triplet masses not too heavy in order to have observable effects, we can estimate from Eq.\  (\ref{eq:vD}) the size of the $\lambda_6$ coupling:  
\begin{equation}
\lambda_6 \ \simeq \ 3.3\times 10^{-12} \left(\frac{M_\Delta}{100~{\rm GeV}}\right)\left(\frac{v_\Delta}{1~{\rm eV}}\right) \, .
\label{eq:lam6}
\end{equation}
This means that $\lambda_6$ will be negligible for what follows, in particular for the correction to the Higgs mass to be discussed in Section~\ref{sec:nat}. 

\subsection{Masses}
Expanding the neutral scalar fields $\phi^0$ and $\delta^0$ around their VEVs,  
\begin{align}
\phi^0 \ = \ \frac{1}{\sqrt{2}}(v+\phi+i\chi),  \qquad 
\delta^0  \ = \ \frac{1}{\sqrt{2}}(v_{\Delta}+\delta+i\eta)\,,
\end{align}
we obtain 
\begin{equation}
\Phi \ = \ \left(
	\begin{matrix}
	\phi^+ \\
	\frac{1}{\sqrt{2}} (v+\phi+i\chi)
	\end{matrix}
\right),
\qquad
\Delta \ = \ \left(
\begin{matrix}
\frac{\delta^+}{\sqrt{2}} & \delta^{++} \\
\frac{1}{\sqrt{2}}(v_{\Delta} + \delta + i \eta) & -\frac{\delta^+}{\sqrt{2}}
\end{matrix}
\right),
\label{eq:neutral}
\end{equation}
which leads to 10 real scalar fields (4 from $\Phi$ and 6 from $\Delta$). Three of them are massless Goldstone bosons $G^{\pm}$, $G^0$, which give masses to the EW gauge bosons $W^{\pm}$, $Z$. So there remain seven physical massive eigenstates, denoted here by $h$, $H^0$, $A^0$, $H^{\pm}$, $H^{\pm \pm}$.  With small $v_\Delta \ll v$, the mixing 
among the doublet and triplet scalars are small. Neglecting also the fine-tuned possibility that the two CP-even scalars are 
degenerate in mass, the lightest physical scalar can be identified as the SM Higgs with mass eigenvalue $m_h^2 \simeq \lambda v^2$, essentially unchanged from the SM case, while the remaining masses are given by 
\begin{align}
m_{H^{\pm \pm}}^2 & \ \simeq \  M_{\Delta}^2+\frac{1}{2}(\lambda_4 + \lambda_5)v^2\,, \label{Eq:mH++}\\
m_{H^{\pm}}^2 & \ \simeq \ M_{\Delta}^2 + \frac{1}{2} \lambda_4 v^2\,, \label{Eq:mH+}\\
m_{A^0, H^0}^2 & \ \simeq \ M_{\Delta}^2 + \frac{1}{2} (\lambda_4 - \lambda_5)v^2\,. \label{Eq:mAmH}
\end{align}
Note that the splitting between the dominantly triplet scalar masses is proportional to $\lambda_5 v^2$. In the case $M_{\Delta}^2\gg v^2$, all of them would be degenerate with mass $M_{\Delta}$. However, for triplet masses close to the EW scale $M_{\Delta}^2\sim v^2$, the mass splitting could be noticeable. For example, for a mass of the single charged triplet component $m_{H^{\pm}}^2=400~\mathrm{GeV}$ and a coupling $\lambda_5=0.5$, the splitting would be $\sim 20~\mathrm{GeV}$.

\section{\label{sec:Constraints}Constraints on the Model}

\subsection{\label{sec:nat}Correction to the Higgs mass and Naturalness}
\begin{figure}[t!]
\includegraphics[width=3.8cm]{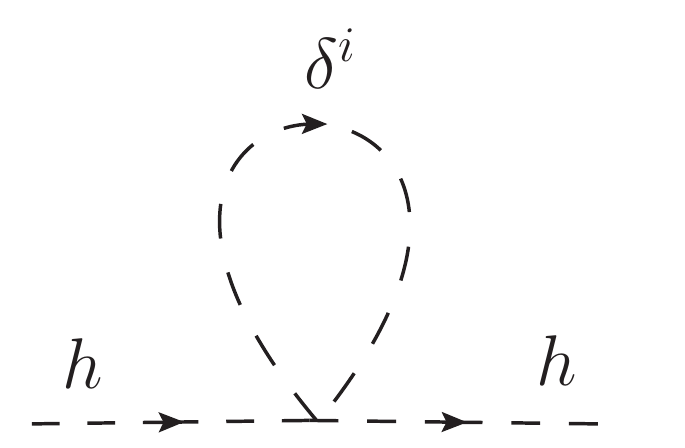}\quad
\includegraphics[width=4.5cm]{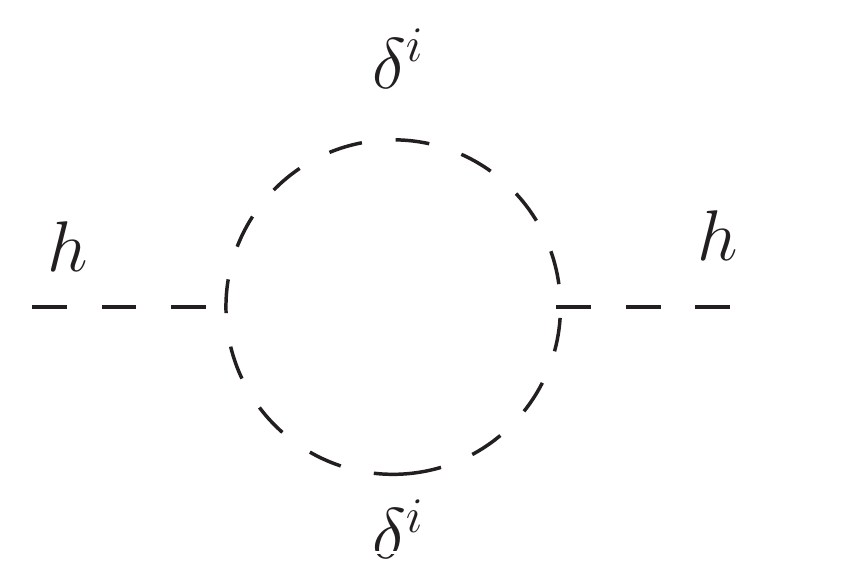}\quad
\includegraphics[width=4.5cm]{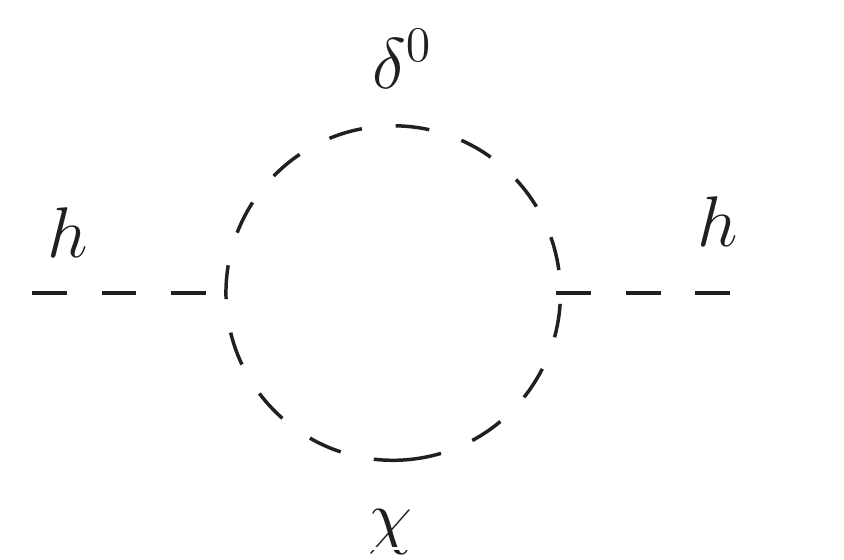}
\caption{One-loop corrections to the Higgs mass in the type II seesaw
  model. Here $\delta^i$ stands for the $\delta^\pm$ and
  $\delta^{\pm\pm}$ fields, and $\chi$ for the imaginary part of the neutral doublet field [cf.~Eq.~\eqref{eq:neutral}].} 
\label{fig:mass}
\end{figure}
From the scalar potential~\eqref{Eq:ScalarPotentialSWII} of the type II seesaw model, one can see that  the new heavy scalar triplet directly couples to the SM Higgs doublet through the $\lambda_4,\lambda_5$ and $\Lambda_6$ terms. These couplings give rise to one-loop corrections involving the heavy scalar triplets to the bare Higgs mass, as shown in Figure~\ref{fig:mass}. 
Applying the Feynman gauge, using $d$-dimensional regularization and the $\overline{\mbox{MS}}$-scheme to remove the infinities that appear in the momentum integrals through counterterms, we obtain
\begin{equation}
\delta m_h^2 \ = \ -\frac{3}{16 \pi^2} \left( \lambda_4 + \frac{\vert
    \lambda_6 \vert ^2}{2}  \right) M_{\Delta}^2 \left[ 1
  +\ln\left(\frac{\mu^2_R}{M_{\Delta}^2}\right) \right] , 
\label{Eq:SWII-HiggsMassCorrection}
\end{equation}
where $\mu_R$ is the regularization scale. 
Here the dimensionless coupling $\lambda_6$ is defined as $\lambda_6\equiv\Lambda_6/M_{\Delta}$ 
and we have neglected terms proportional to $v^2$. For definiteness, we will choose $\mu_R=M_\Delta$ and thus set the term in squared brackets to $1$.  Note that $\lambda_5$ does not appear in the final result, though diagrams involving it and contributing to the mass correction exist. This happens \cite{Abada:2007ux} because  
of the $\Phi^{\dagger} [\Delta^{\dagger},\Delta]\Phi$ structure of the term in the potential. It uses the Higgs fields as $SU(2)$ triplet, while whatever is proportional to 
Higgs mass must be a singlet. From \eqref{Eq:SWII-HiggsMassCorrection}, neglecting the $\lambda_6$ contribution, since it is much smaller [cf.~Eq.~\eqref{eq:lam6}], we find that 
\begin{equation}\label{eq:res2}
\frac{\delta m_h^2}{m_h^2} \ \simeq \
 -0.01 \lambda_4 \left(\frac{M_\Delta}{100\, \rm GeV} \right)^2 \,.
\end{equation} 
Thus, TeV-scale triplet masses pose a problem for naturalness, unless the quartic coupling $\lambda_4\ll 1$. This is in 
contrast to naturalness in case of the type I seesaw, where the correction $\delta m_h^2$ 
is of order $m_\nu  M_N^3 /v^2$, and limits of order $M_N \ls 10^7$ GeV on the right-handed neutrino mass arise~\cite{Bambhaniya:2016rbb}.

\subsection{\label{sec:UP}Stability, Unitarity and Perturbativity}

The necessary and sufficient conditions which ensure the potential~\eqref{Eq:ScalarPotentialSWII} of the type II seesaw model is bounded from below have been studied in Refs.~\cite{Bonilla, Arhrib, Das:2016bir}. 
Taking into account all field directions, they can be written as follows:  
\begin{subequations}
\begin{align}
&\lambda \ \geq \ 0\,, \label{Eq:StabCond1}\\ 
&\lambda_1 \ \geq \  0\,, \label{Eq:StabCond2}\\
&2\lambda_1+\lambda_2 \ \geq \ 0\,, \label{Eq:StabCond3}\\
&\lambda_4+\lambda_5+\sqrt{\lambda \lambda_1} \ \geq \ 0\,, \label{Eq:StabCond4} \\
&\lambda_4-\lambda_5+\sqrt{\lambda\lambda_1} \ \geq \ 0\, , \label{Eq:StabCond5}\\
&2|\lambda_5| \sqrt{\lambda_1} + \lambda_2\sqrt{\lambda} \ \geq \ 0 \quad \mathrm{or} \quad\lambda_4+\sqrt{\left(\lambda \lambda_2 + 2 \lambda_5^2\right)\left(\lambda_1/\lambda_2+1/2\right)} \ \geq \ 0\,. \label{Eq:disjunction}
\end{align}
\end{subequations}
Two further stability conditions, recently given in Ref.~\cite{Bonilla}, are 
\begin{subequations}
\begin{align}
\lambda_4+\lambda_5+\sqrt{\lambda\left(\lambda_1+\frac{\lambda_2}{2}\right)} & \ \geq \ 0\,, \\
\lambda_4-\lambda_5+\sqrt{\lambda\left(\lambda_1+\frac{\lambda_2}{2}\right)} &\  \geq \ 0\,.
\end{align}
\end{subequations}
Those conditions are sufficient but not necessary, and have to be substituted by the disjunction 
Eq.~\eqref{Eq:disjunction}, in which one of the two conditions must 
be satisfied in order the have a bounded from below potential.

In addition, constraints on the scalar potential parameters can be obtained by demanding tree-level unitarity to be preserved in a variety of scattering processes: scalar-scalar scattering, gauge-boson-gauge-boson scattering, and scalar-gauge-boson scattering. These have been studied in the type II seesaw model e.g.\ in Ref.~\cite{Arhrib} (see also references therein). 
Demanding the tree level unitarity to be preserved for different elastic scattering 
processes the following constraints are obtained: 
\begin{subequations}
\begin{align}
& \lambda \ \leq \ \frac{8}{3} \pi\,, \label{Eq:Scat1}\\ 
&\lambda_1-\lambda_2 \ \leq \ 8\pi\,, \\
&4\lambda_1+\lambda_2 \ \leq \ 8\pi\,, \\
&2\lambda_1+3\lambda_2 \ \leq \ 16\pi\,, \\
& \vert\lambda_5\vert \ \leq \ \frac{1}{2}\mathrm{min}\left[ \sqrt{(\lambda \pm 8\pi)(\lambda_1-\lambda_2\pm 8\pi)}\right]\,, \\
& \vert \lambda_4\vert \ \leq \ \frac{1}{\sqrt{2}} \sqrt{\left( \lambda - \frac{8}{3}\pi\right)(4\lambda_1+\lambda_2-8\pi)}\label{Eq:Scat6}\,.
\end{align}
\end{subequations}
Finally, perturbativity implies that all dimensionless parameters are smaller than $4\pi$.

\subsection{\label{sec:direct}Direct Constraints from the LHC}

Direct limits on the triplet masses have been derived from collider searches of multi-lepton final states. A typical process at hadron colliders is the Drell-Yan pair production of  
$\delta^{++} \delta^{--}$ through $s$-channel $\gamma/Z$ exchange, with subsequent decay of $\delta^{\pm\pm}$ into charged lepton pairs, which is the dominant decay mode for $v_\Delta\lesssim 0.1$ MeV~\cite{Perez:2008ha, Melfo:2011nx}. One can also have the associated production of $\delta^{\pm\pm}\delta^\mp$ through $s$-channel $W^\pm$ exchange, followed by the $\delta^{\pm\pm}$ decay to charged lepton pair and $\delta^\mp$ decay to a charged lepton and neutrino final state. In both cases, the limits on the triplet mass will depend on the 
relative branching ratios to different final state charged lepton flavors. For instance, using the $\sqrt s= 13$ TeV LHC  data, both ATLAS and CMS experiments have set a lower limit on the doubly-charged scalar masses from about 400 GeV to 800 GeV, depending on the final state lepton flavor~\cite{ATLAS:2016pbt, CMS:2017pet}.   
We will assume here the absolute lower limit of $m_{\Delta}>400~\mathrm{GeV}$ valid for all flavors. In the future, these limits can be 
improved by about a factor 2 with the high-luminosity (HL) LHC, and by a factor 10 with a 100 TeV $pp$ collider \cite{Dev:2016dja,Mitra:2016wpr}.

There is also a constraint from EW precision data (the oblique parameter $T$) 
on the mass difference between the doubly and singly charged triplet scalars~\cite{Lavoura:1993nq}. 
It constrains this splitting to 
\begin{equation}
|\Delta M| \ \equiv \ |m_{H^{++}}-m_{H^+}| \ \lesssim \ 40~\mathrm{GeV}\,,
\label{Eq:EWPDConstrain}
\end{equation} 
almost independently of the double charged Higgs mass \cite{Chun, Melfo:2011nx}. 
From Eqs.\ (\ref{Eq:mH++})-(\ref{Eq:mAmH}) we see that the mass splitting among the 
triplet components is induced by the $\lambda_5$-coupling:
\begin{equation}
m^2_{H^{++}}-m^2_{H^+} \ = \ \frac{1}{2}\lambda_5 v^2\,.
\label{Eq:ChargedHiggsMassSplitting}
\end{equation}
Taking typical lower limits from direct searches, the constraint becomes 
\begin{equation}\label{eq:lam5}
-1.1 \ \lesssim \ \lambda_5 \ \lesssim \ 1\,. 
\end{equation}

\subsection{\label{sec:LFV}Lepton Flavor Violation}

The seven physical bosons introduced in the type II seesaw model contribute to many LFV processes. We will see  that the low scale seesaw case, with $400~\mathrm{GeV}<M_{\Delta}<1~\mathrm{TeV}$, is severely constrained by experiments searching for LFV, which have set stringent bounds on the branching ratio of these processes.

The singly and double charged Higgs scalars $H^+$ and $H^{++}$ have, in general, different masses, $m_{H^+}$ and $m_{H^{++}}$, with a splitting of the squared masses of $\tfrac{1}{2}\lambda_5 v^2$ [see Eqs.\ (\ref{Eq:mH++}), (\ref{Eq:mH+})]. Since the sign of $\lambda_5$ is not known a priori, both $m_{H^{+}}>m_{H^{++}}$ and $m_{H^{+}}<m_{H^{++}}$ are possible. For values of the triplet mass scale much larger than the EW scale this splitting is negligible, whereas for values close to the EW scale the splitting could be of few $\mathrm{GeV}$ and therefore noticeable. However, even in the case of a low-scale seesaw, the impact of the splitting in the LFV branching ratios is almost not perceptible, and therefore not relevant in our study. Thus, the mass difference will be neglected in the following and we will consider $m_{H^+} \simeq m_{H^{++}} = M_{\Delta}$. 

The branching ratio for $\mu \rightarrow e \gamma$ is given by \cite{Akeroyd,Dinh}
\begin{equation}
\mathrm{BR} (\mu \rightarrow e \gamma) \ = \ \frac{27 \alpha _{\mathrm{em}}}{64 \pi G_F^2 M_{\Delta}^4} \vert (Y_{\Delta}^\dagger Y_{\Delta})_{e \mu} \vert ^2 \, \mathrm{BR} (\mu \rightarrow e \bar{\nu} \nu)\,,
\label{Eq:BRmuegamma}
\end{equation}
where $\alpha_{\mathrm{em}} \equiv q_e^2/4 \pi = 1/137$ is the fine structure constant, $G_F=1.17 \times 10^{-5} ~\mathrm{GeV}^{-2}$ is the Fermi constant and $\mathrm{BR} (\mu \rightarrow e \bar{\nu} \nu) \simeq 100\% $ \cite{Olive:2016xmw}. The $\mu\to e\gamma$ process provides the strongest constraint on the triplet parameters, and it cannot be 
evaded because $(Y_{\Delta}^\dagger Y_{\Delta})_{e \mu}$ cannot vanish \cite{Chakrabortty}. Similar 
expressions as \eqref{Eq:BRmuegamma} exist for $\tau \rightarrow \mu \gamma$ and 
$\tau \rightarrow e \gamma$. Formulas for the decays 
$\mu \rightarrow 3e$, $\tau \rightarrow \bar{l}_i l_j l_k$ and $\mu$--$e$ conversion in 
nuclei can be found in Refs.\ \cite{Akeroyd,Dinh} as well. The latter are subject to strong variation, 
because they depend on individual elements of $Y_\Delta$, which 
can vanish \cite{Merle-Rodejohann,Grimus}.  Crucial is that 
$(Y_{\Delta}^\dagger Y_{\Delta})_{e \mu}$ 
is essentially fixed by neutrino oscillation data
\cite{Chakrabortty}. 
Looking at \eqref{Eq:BRmuegamma}, the branching ratio is then proportional to $1/(v_\Delta M_\Delta)^4$, which implies a lower limit on the product 
$v_\Delta M_\Delta$. This is the most stringent LFV constraint in the type II seesaw model. 
To be a bit more definitive, the above branching ratio depends on 
\begin{equation}
|(M_\nu^\dagger M_\nu)_{e\mu}|^2 \ \simeq \ (\Delta m^2_{32})^2  \sin^2\theta_{23} \, |U_{e3}|^2 + 
\Delta m^2_{21}  \, \Delta m^2_{32} \cos \delta \cos \theta_{23} \, \sin \theta_{23} \, \sin2\theta_{12} |U_{e3}| \,. 
\label{eq:lfv}
\end{equation}
The best-fit value is $5.4 \times 10^{-8}$ eV$^4$, and the $3\sigma$ range is 
$(0.30 - 1.2) \times 10^{-7}$ eV$^4$, for the normal mass ordering. What is of interest here are 
limits on the triplet mass from LFV. Since $M_\Delta$ appears with the fourth power in the branching ratios, the full range 
of $|(M_\nu^\dagger M_\nu)_{e\mu}|^2$ is not important.

Table~\ref{Table:LFVBounds} summarizes the experimental limits on the branching ratios for the different LFV processes and the corresponding constraints on the various combinations of the Yukawa coupling matrix elements of the leptons to the scalar triplet. These upper bounds on the Yukawa couplings can be used to set lower bounds on the VEV of the scalar triplet $v_{\Delta}$ for a given triplet mass $M_{\Delta}$ using Eq.~(\ref{Eq:MassMatrix}). In general, the prediction depends on the type of hierarchy of the neutrino mass spectrum: normal hierarchy (NH) or inverted hierarchy (IH). For a given spectrum, it can depend on the Majorana and Dirac phases, as well as on the value of the lightest neutrino mass, $m_{\nu_{\mathrm{min}}}$. 

For illustration purposes, we have computed the lower limit of the product $v_{\Delta}M_{\Delta}$ for NH and IH considering $m_{\nu_{\mathrm{min}}}=0$ and $m_{\nu_{\mathrm{min}}}=0.2~\mathrm{eV}$, which satisfy the current upper limits set by  neutrinoless double beta decay experiments and cosmology. We have also used the best-fit values of the neutrino oscillation parameters and considered the cases of zero and non-zero Majorana phases, with $(\alpha_1=\pi/3,\ \alpha_2=\pi/2)$ in the second case, corresponding to no and large cancellations in the elements of $Y_\Delta$, respectively. 
The results are 
summarized in Table~\ref{Table:vMBounds}. Note that taking the lightest neutrino mass to zero gives the least restrictive lower limit on $v_{\Delta}M_{\Delta}$, since it corresponds 
to the smallest values for the Yukawa couplings.

From Table~\ref{Table:vMBounds}, it can be observed that the most stringent bounds come from the $\mu \rightarrow e \gamma$ and $\mu \rightarrow 3e$ decays; $\mathrm{BR}(\mu \rightarrow e \gamma)$ is independent of the Majorana phases [cf.~Eq.~\eqref{eq:lfv}], since it is proportional to $|(Y_{\Delta}^{\dagger}Y_{\Delta})_{e\mu}|$, and the diagonal matrix containing the Majorana phases, $\mathrm{diag}(1,e^{i\alpha_1},e^{i\alpha_2})$, cancels in the product $(Y_{\Delta}^{\dagger}Y_{\Delta})$. Furthermore, it is also independent of the absolute neutrino mass. The $\mathrm{BR}(\mu \rightarrow 3e)$ depends on the individual entries of the Yukawa matrix $\vert (Y_{\Delta})_{\mu e} \vert$ and  $\vert (Y_{\Delta})_{ee} \vert$, which can vanish for specific values of the Majorana phases \cite{Merle-Rodejohann,Grimus}. In view of this dependence of the bounds on the value of the Majorana phases, we will consider in what follows mostly the limits set by the 
$\mu \rightarrow e\gamma$ process.

\begin{table}[t!]
\begin{center}
\begin{tabular}{c | c | c | c  }
\hline\hline
 Process & \makecell{Experimental limit \\on BR} & Constraint on & Bound $\displaystyle \times \left(\frac{M_{\Delta}}{100~\mathrm{GeV}}\right)^2 $ \\ \hline\hline
$\mu \rightarrow e \gamma$ & $<4.2 \times 10^{-13}$ \cite{MEG} 
& $\vert (Y_{\Delta}^\dagger Y_{\Delta})_{e \mu} \vert $ & $<2.4 \times 10^{-6}$ \\ 
$\mu \rightarrow 3e$ & $< 1.0 \times 10^{-12}$ \cite{SINDRUM} & $\vert (Y_{\Delta})_{\mu e} \vert  \vert (Y_{\Delta})_{ee} \vert $ & $<2.3 \times 10^{-7}$ \\ \hline  
$\tau \rightarrow e \gamma$ & $<3.3 \times 10^{-8}$ \cite{Olive:2016xmw} & $\vert (Y_{\Delta}^\dagger Y_{\Delta})_{e \tau} \vert $ & $<1.6 \times 10^{-3}$ \\ 
$\tau \rightarrow \mu \gamma$ & $<4.4 \times 10^{-8}$ \cite{Olive:2016xmw} & $\vert (Y_{\Delta}^\dagger Y_{\Delta})_{\mu \tau} \vert $ & $<1.9 \times 10^{-3}$ \\ 
$ \tau \rightarrow e^+ e^- e^- $ & $<2.7 \times 10^{-8}$ \cite{Olive:2016xmw} & $\vert (Y_{\Delta})_{\tau e} \vert  \vert (Y_{\Delta})_{ee} \vert $ & $<9.2 \times 10^{-5}$ \\ 
$ \tau \rightarrow \mu^+ \mu^- e^- $ & $<2.7 \times 10^{-8}$ \cite{Olive:2016xmw} & $\vert (Y_{\Delta})_{\tau \mu} \vert  \vert (Y_{\Delta})_{\mu e} \vert $ & $<6.5 \times 10^{-5}$ \\ 
$ \tau \rightarrow e^+ \mu^- \mu^- $ & $<1.7 \times 10^{-8}$ \cite{Olive:2016xmw} & $\vert (Y_{\Delta})_{\tau e} \vert  \vert (Y_{\Delta})_{\mu \mu} \vert $ & $<7.3 \times 10^{-5}$ \\ 
$ \tau \rightarrow e^+ e^- \mu^- $ & $<1.8 \times 10^{-8}$ \cite{Olive:2016xmw} & $\vert (Y_{\Delta})_{\tau e} \vert  \vert (Y_{\Delta})_{\mu e} \vert $ & $<5.3 \times 10^{-5}$ \\ 
$ \tau \rightarrow \mu^+ e^- e^- $ & $<1.5 \times 10^{-8}$ \cite{Olive:2016xmw} & $\vert (Y_{\Delta})_{\tau \mu} \vert  \vert (Y_{\Delta})_{ee} \vert $ & $<6.9 \times 10^{-5}$ \\ 
$ \tau \rightarrow \mu^+ \mu^- \mu^- $ & $<2.1 \times 10^{-8}$ \cite{Olive:2016xmw} & $\vert (Y_{\Delta})_{\tau \mu} \vert  \vert (Y_{\Delta})_{\mu \mu} \vert $ & $<8.1 \times 10^{-5}$\\ \hline\hline
\end{tabular}
\end{center}
\caption{Experimental limits on the branching ratios of different LFV processes and the corresponding bounds on different combinations of $Y_{\Delta}$  in the type II seesaw model.}
\label{Table:LFVBounds}
\end{table}
\begin{table}[t!]
\begin{center}
\def\arraystretch{1.2}
\begin{tabular}{c | c  c | c  c }
\hline \hline
\multirow{3}{*}{Process} & \multicolumn{4}{c}{Lower limit on $\displaystyle\left(\frac{v_{\Delta}}{\mathrm{eV}}\right)\left(\frac{ M_{\Delta}}{100~\mathrm{GeV}}\right)$} \\ 
\cline{2-5}
& \multicolumn{2}{c} {NH} & \multicolumn{2}{|c} {IH} \\ 
\cline{2-5}
& $m_1=0~\mathrm{eV}$ & $m_1=0.2 ~\mathrm{eV}$ & $m_3=0~\mathrm{eV}$ & $m_3=0.2 ~\mathrm{eV}$ \\ \hline \hline
$\mu \rightarrow e \gamma$ & $>6.9\ (6.9)$ & $>6.9\ (6.9)$ & $>7.4\ (7.4)$ & $>7.4\ (7.4)$\\ 
$\mu \rightarrow 3e$ & $>4.5\ (3.5)$ & $>119.6\ (167.6)$ & $>23.6\ (41.5)$ & $>127.0\ (157.5)$\\ \hline  \hline
$\tau \rightarrow e \gamma$ & $>0.30\ (0.30)$ & $>0.30\ (0.30)$ & $>0.30\ (0.30)$ & $>0.30\ (0.30)$\\ 
$\tau \rightarrow \mu \gamma$ & $>0.60\ (0.57)$ & $>0.57\ (0.57)$ & $>0.57\ (0.57)$ & $>0.57\ (0.57)$\\ 
$ \tau \rightarrow e^+ e^- e^- $ & $>0.28\ (0.37)$ & $>6.42\ (8.82)$& $>1.14\ (2.14)$& $>6.0\ (9.60)$\\ 
$ \tau \rightarrow \mu^+ \mu^- e^- $ & $>0.90\ (0.57)$ & $>0.69\ (9.73) $ & $>1.00\ (1.72)$ & $>1.36\ (8.70)$\\ 
$ \tau \rightarrow e^+ \mu^- \mu^- $ & $>1.11\ (1.14)$ & $>7.33\ (10.1)$ & $>0.84\ (1.54)$ & $>6.70\ (11.5)$\\ 
$ \tau \rightarrow e^+ e^- \mu^- $ & $>0.55\ (0.37)$ & $>3.58\ (10.8)$ & $>0.50\ (2.68)$ & $>3.42\ (10.8)$\\ 
$ \tau \rightarrow \mu^+ e^- e^- $ & $>0.60\ (0.72)$ & $>1.58\ (10.2)$ & $>2.96\ (1.77)$ & $>3.05\ (9.96)$\\ 
$ \tau \rightarrow \mu^+ \mu^- \mu^- $ & $>1.82\ (1.85)$ & $>1.47\ (9.55)$ & $>1.79\ (1.04)$ & $>2.80\ (9.80)$\\ \hline \hline
\end{tabular}
\end{center}
\caption{Lower limit on the product $v_{\Delta}M_{\Delta}$ obtained from the experimental bounds on the branching ratio of different LFV processes (Table~\ref{Table:LFVBounds}), for NH and IH, calculated using the best fit values of the neutrino oscillation data \cite{Esteban:2016qun}, with Majorana phases $\alpha_1=0$, $\alpha_2=0$ ($\alpha_1=\pi/3$, $\alpha_2=\pi/2$) and assuming the lightest neutrino mass to be $0~\mathrm{eV}$ and $0.2~\mathrm{eV}$.}
\label{Table:vMBounds}
\end{table}

\section{\label{sec:res}Results and discussion}
\subsection{\label{sec:allowed}Allowed Parameter Space}

In this section, we will use the constraints from vacuum stability and unitarity of scattering processes, Eqs.~(\ref{Eq:StabCond1})-(\ref{Eq:disjunction}) and Eqs.~(\ref{Eq:Scat1})-(\ref{Eq:Scat6}) respectively, and by imposing them to be fulfilled up to the Planck scale, restrict the parameter space in the type II seesaw model.

To ensure that the vacuum stability and the unitary conditions presented above 
are fulfilled up to the Planck scale, it is necessary to study their 
renormalization group equations (RGEs). Depending on whether the renormalization scale 
$\mu$ is below or above the new energy scale, determined by $M_{\Delta}$, the RG running 
will be different, because below the scale of $M_\Delta$, the scalar triplet can be integrated out and we are effectively left with the SM. 
We will employ two-loop RGEs for the SM couplings and one-loop RGEs for the 
new couplings associated with the type II seesaw scenario, following the procedure 
specified in Ref.~\cite{Dev:2013ff}. The one-loop RGEs for the various scalar couplings 
$\lambda_{1,2,4,5}$ ($\lambda_6$ is decoupled at one-loop level) 
in the type II seesaw model are given in Refs.~\cite{Dev:2013ff, Chao:2006ye, Schmidt:2007nq, Gogoladze:2008gf} and we do not write them here explicitly. The equations 
depend on the quartic couplings $\lambda_{1,2,4,5}$, the gauge couplings $g$ and $g'$,  
and the Yukawa couplings $Y_\Delta^\dagger Y_\Delta$. Below the scale $M_\Delta$, the effective quartic SM Higgs coupling is shifted down to 
$\lambda \to \lambda - \lambda_6^2$, which in our case of tiny $\lambda_6$ [cf.~Eq.~\eqref{eq:lam6}] is not of importance. 
The running of $\lambda$ also depends on the top quark Yukawa coupling, which is the only fermion Yukawa we have considered in the RGEs.  
The boundary conditions can be determined from the one-loop matching conditions, described in Ref.~\cite{Dev:2013ff}. 
If the energy scale is larger than $M_\Delta$, the gauge and $\lambda$ couplings receive additional contributions, 
in particular the beta-function of $\lambda$ receives a positive contribution from $\lambda_4^2$ and $\lambda_5^2$, which 
helps to improve the EW vacuum stability. 
We first run the gauge couplings from their values at the $Z$ mass to the top mass, then set the boundary conditions for 
the Higgs quartic and top Yukawa coupling at the top mass, then run them to $M_\Delta$, and from there, run them to the Planck scale with the triplet scalar contributions. 

Using this procedure, we analyze numerically the parameter space in the scalar sector of the type II seesaw model 
which satisfies the vacuum stability, unitarity and perturbativity 
conditions discussed in Section~\ref{sec:UP} up to the Planck scale. 
We also impose the different constraints coming from direct searches and LFV experiment, explained in Section~\ref{sec:direct} and~\ref{sec:LFV} respectively, which fix the allowed values of the triplet mass $M_{\Delta}$ and triplet VEV $v_{\Delta}$, and check that the restriction on $\lambda_5$ from EW precision data [cf.~Eq.~\eqref{eq:lam5}] is fulfilled. 
In addition, we also study which part of the allowed parameter space fulfills the 
naturalness condition (cf.~Section~\ref{sec:nat}), which we take as $|\delta m_h^2|\lesssim m_h^2$, i.e.\ 
that the radiative correction to the squared Higgs mass is, at most, of 
the order of the physical Higgs mass squared, $m_h^2\simeq(125~\mathrm{GeV})^2$.

In particular, we focus on the low scale seesaw, so we only consider the case in which the scale of new physics $M_{\Delta}$ is in the TeV range, $M_{\Delta}\sim (400~\mathrm{GeV}-3~\mathrm{TeV})$, which could be testable at the LHC and future colliders~\cite{Dev:2016dja}. For simplicity, we assume that the mass splitting of the triplet components is negligible, so that $m_{H^{++}}\simeq m_{H^+}\equiv M_{\Delta}$. This would anyway have only small effect on our results. 
We also restrict the study to small values for the triplet VEV 
$v_{\Delta}\sim\mathcal{O}(\mathrm{eV})$, in order to have sizable Yukawa couplings.

To obtain the allowed parameter space, we randomly generate sets of $(\lambda_1,\lambda_2,\lambda_4,\lambda_5)$ for a fixed triplet mass and VEV, which fulfill the perturbativity, vacuum stability and unitarity conditions. We take these as initial values at $\mu=M_{\Delta}$ and solve simultaneously their one-loop RGEs up to the Planck scale. The value of the Higgs quartic coupling $\lambda$ at $\mu=M_{\Delta}$ is obtained by running its SM RGE up to $\mu=M_{\Delta}$. At this energy its RGE is modified to account for the new contributions coming from the interaction with the scalar triplet. During the running of the couplings it is checked that the perturbativity, vacuum stability and unitarity conditions are always satisfied at each intermediate scale, so that only those sets of parameters that satisfy them up to the Planck scale are kept.

\begin{figure}[t]
    \centering
    \subfloat[]
        {\includegraphics[width=0.4\textwidth]{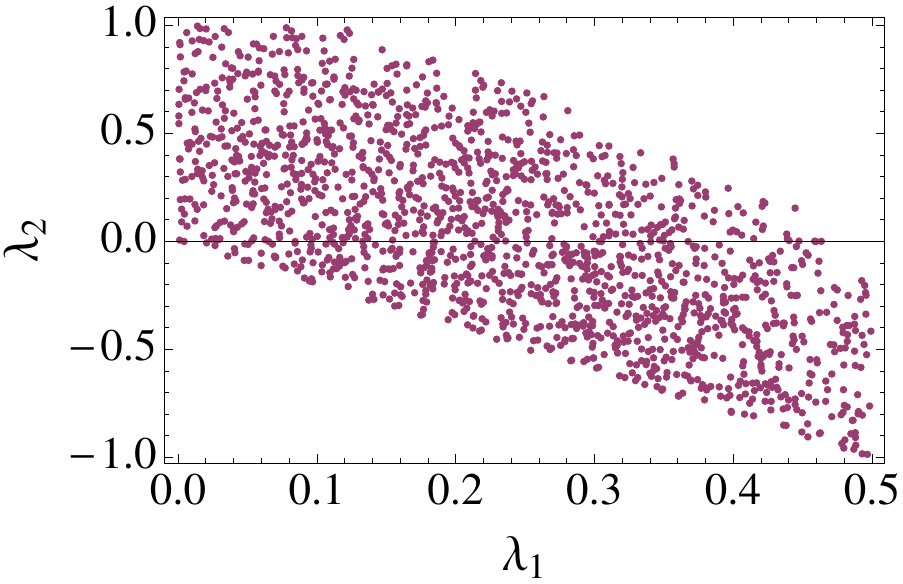}
        \label{Fig:ParamScanNH1a}}
        \qquad
    \subfloat[]
        {\includegraphics[width=0.4\textwidth]{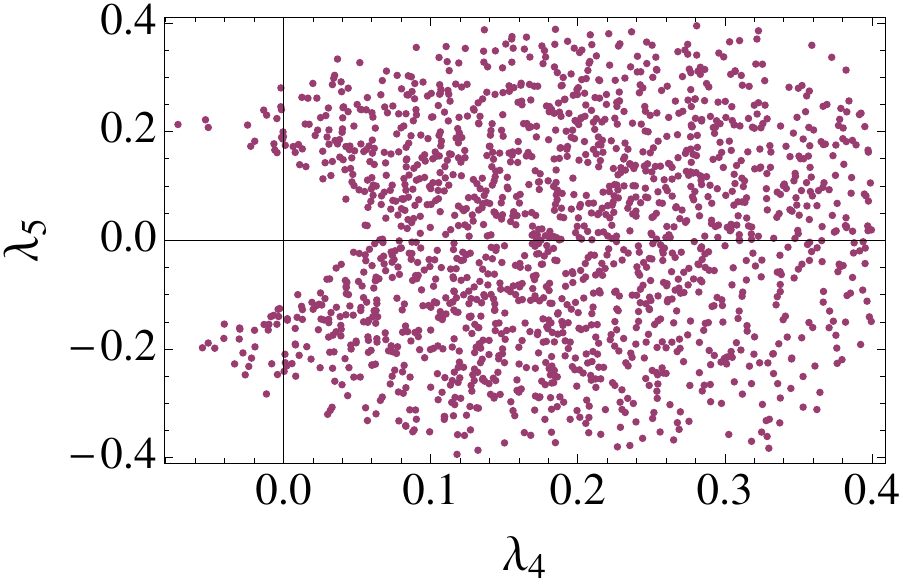}
        \label{Fig:ParamScanNH1b}}
        \\
    \subfloat[]
    {\includegraphics[width=0.42\textwidth]{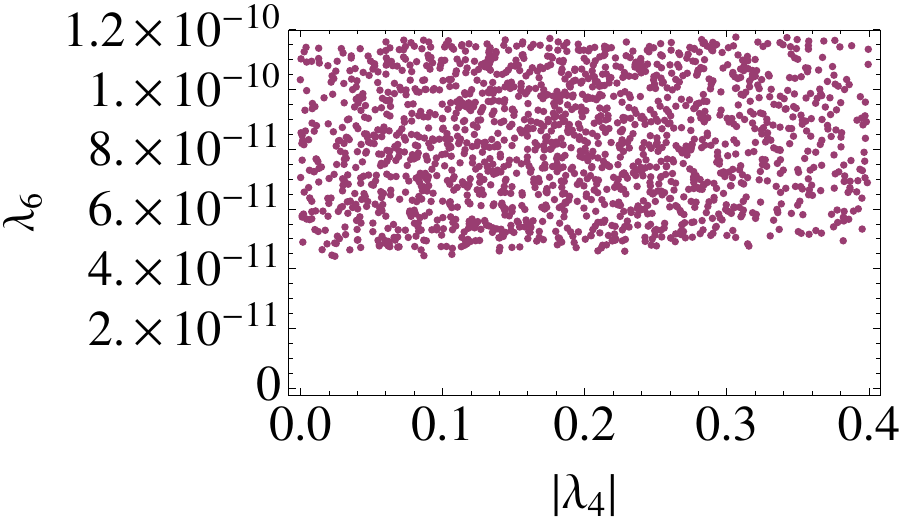}
        \label{Fig:ParamScanNH1c}}
        \qquad
    \subfloat[]
{\includegraphics[width=0.4\textwidth]{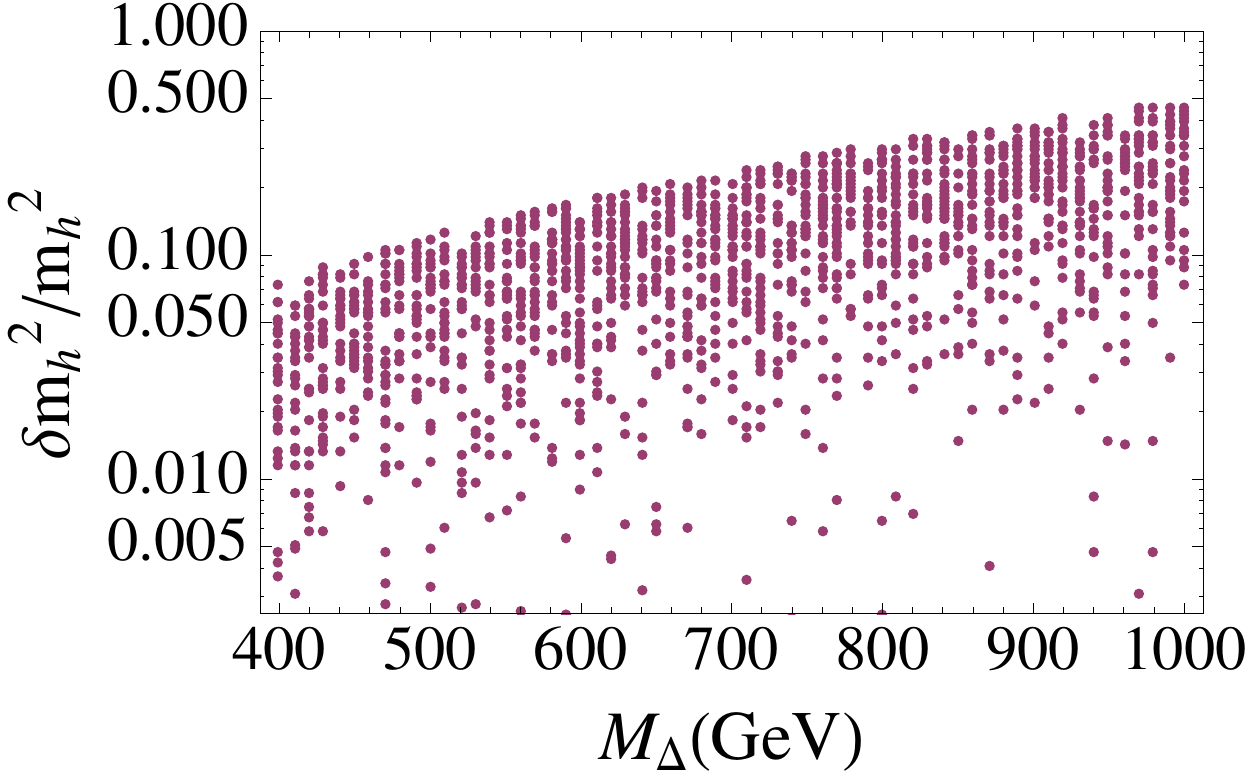}
        \label{Fig:ParamScanNH1d}}
   \caption{Allowed parameter space in the (a) $(\lambda_1,\lambda_2)$ plane, (b) $(\lambda_4,\lambda_5)$ plane, (c) $(|\lambda_4|,\lambda_6)$ plane and (d) $(M_{\Delta}, |\delta m_h^2|/m_h^2)$ plane, for $v_{\Delta}=3.5~\mathrm{eV}$ and $400~\mathrm{GeV}<M_{\Delta}<1~\mathrm{TeV}$ in the type II seesaw model. It has been calculated considering NH and setting $m_{\nu_{\mathrm{min}}}=0$ and the Majorana phases equal to zero. All points satisfy the vacuum stability, unitarity and perturbativity conditions up to Planck scale and the naturalness condition $|\delta m_h^2|\leq m_h^2$ at $\mu=M_{\Delta}$. The values shown correspond to the parameters at $\mu=M_{\Delta}$. }
    \label{Fig:ParamScanNH1}
\end{figure}

\begin{figure}[t]
    \centering
    \subfloat[]
        {\includegraphics[width=0.4\textwidth]{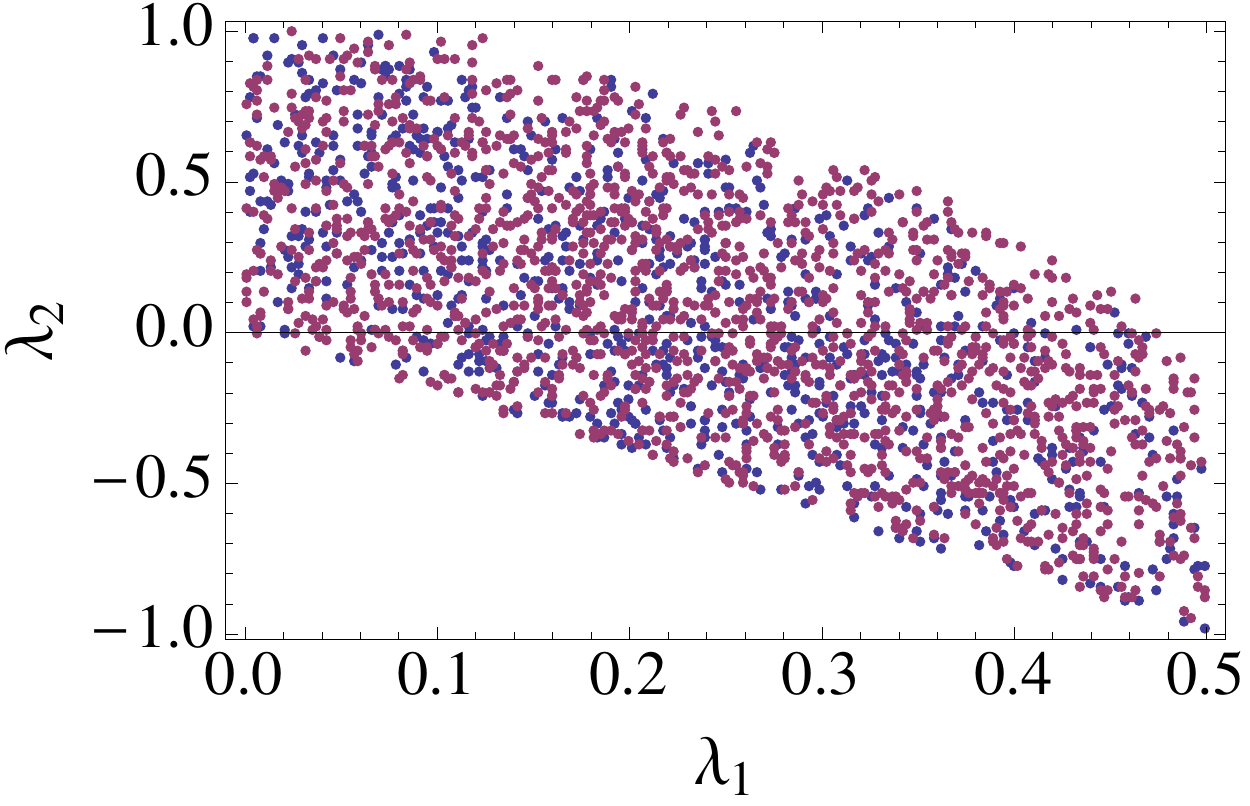}
        \label{Fig:ParamScanNH3a}}
        \qquad
    \subfloat[]
        {\includegraphics[width=0.4\textwidth]{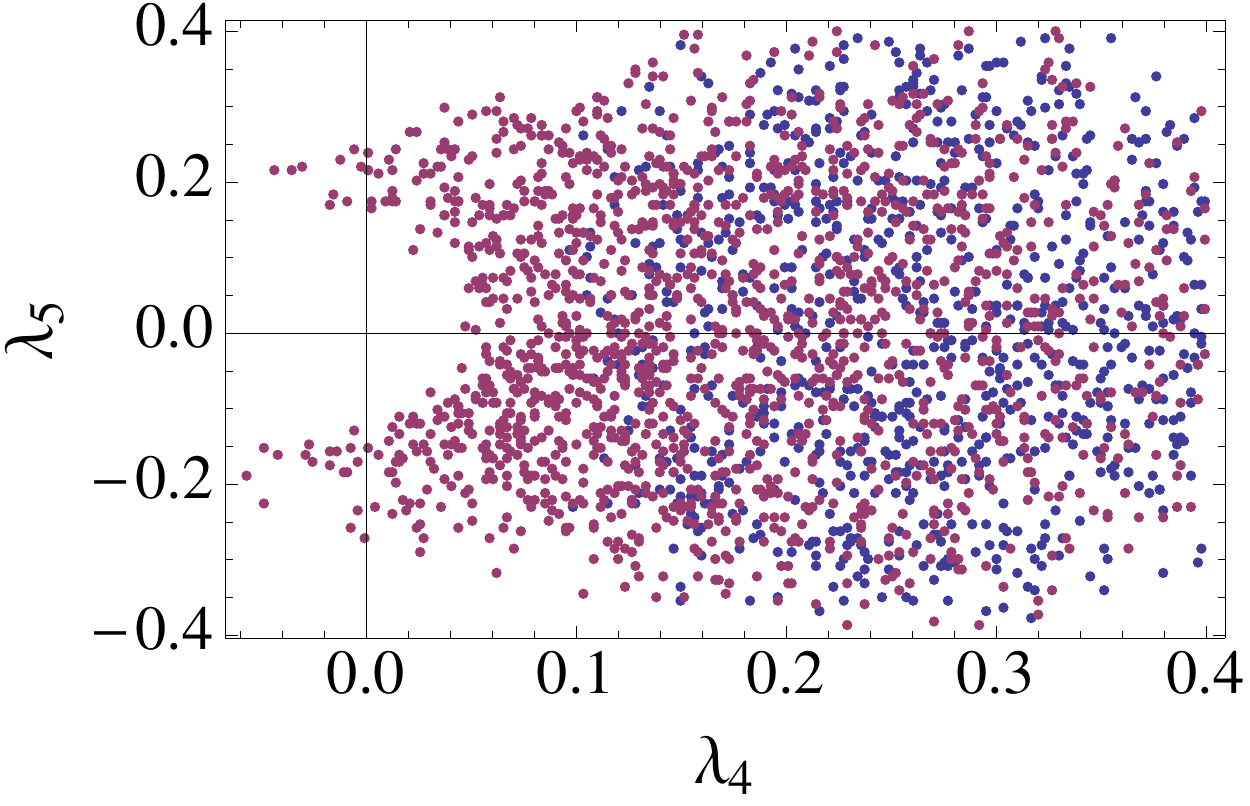}
        \label{Fig:ParamScanNH3b}}
        \\
    \subfloat[]
    {\includegraphics[width=0.42\textwidth]{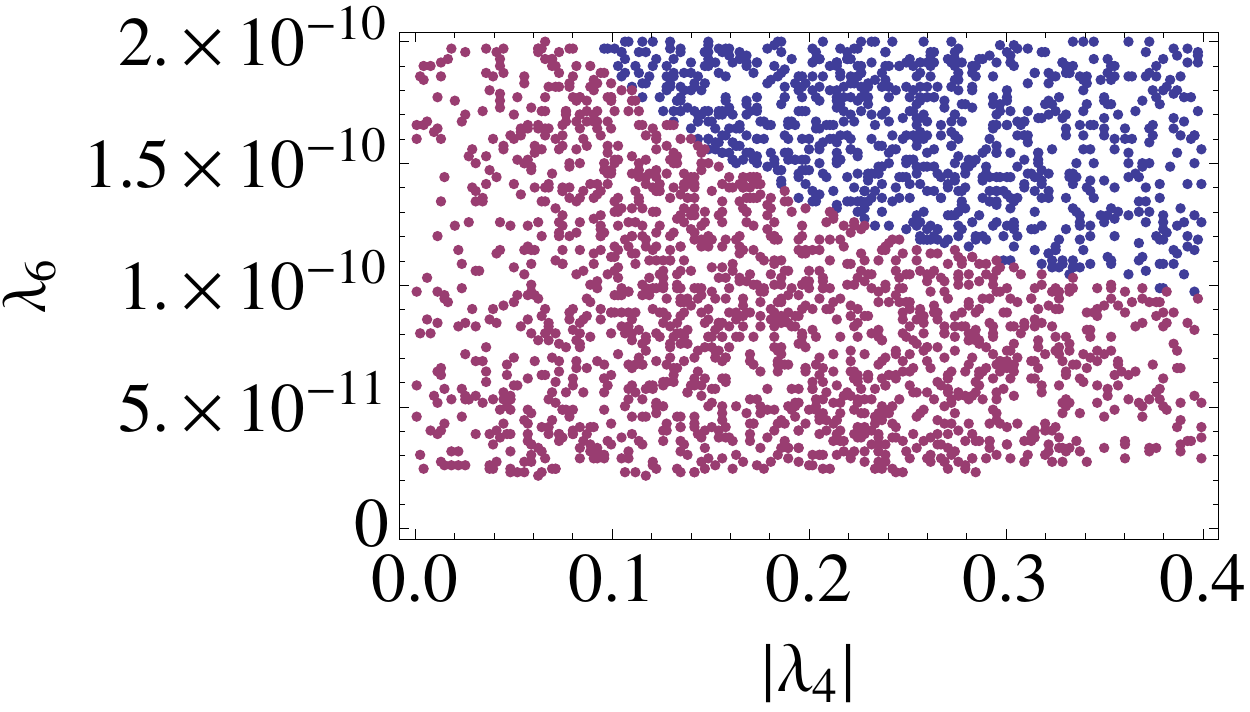}
        \label{Fig:ParamScanNH3c}}
        \qquad
    \subfloat[]
    {\includegraphics[width=0.4\textwidth]{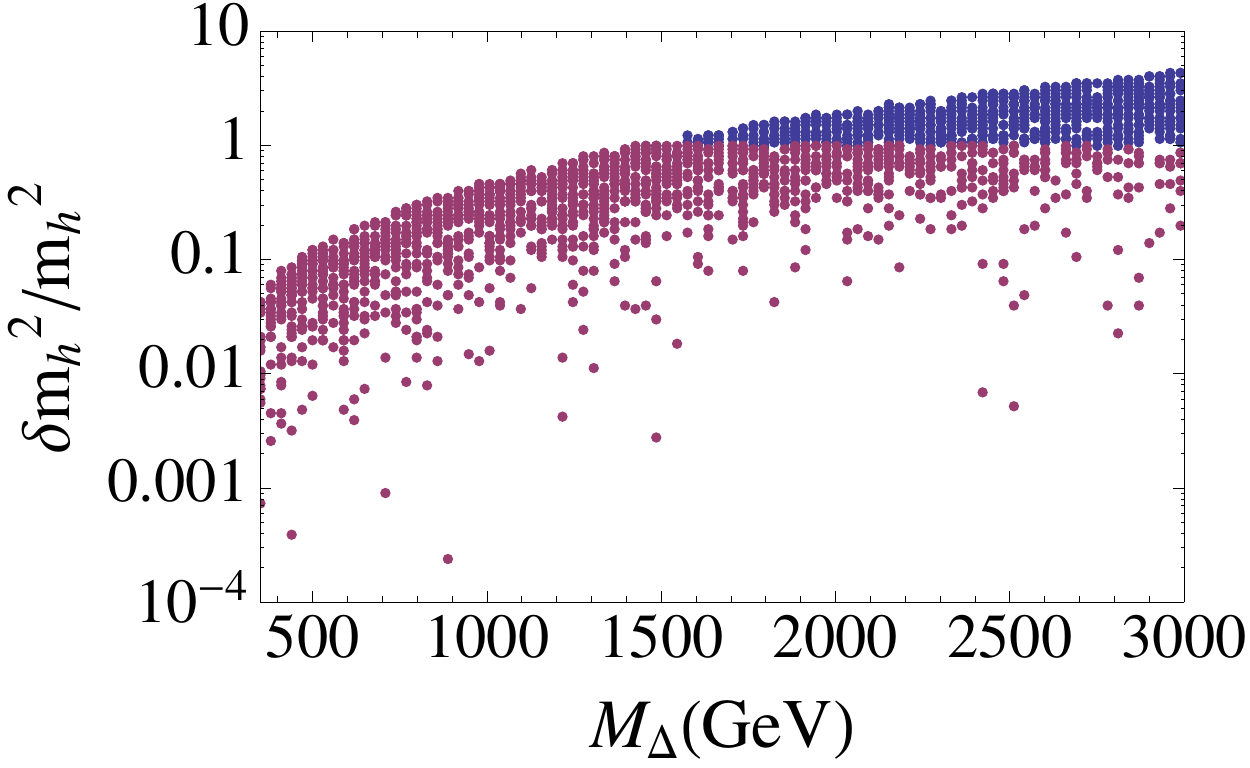}
        \label{Fig:ParamScanNH3d}}
   \caption{Same as Figure~\ref{Fig:ParamScanNH1}, but for $v_{\Delta}=2~\mathrm{eV}$ and $400~\mathrm{GeV}<M_{\Delta}<3~\mathrm{TeV}$. The blue points satisfy the vacuum stability, unitarity and perturbativity conditions up to Planck scale, whereas the pink points also satisfy the naturalness condition $|\delta m_h^2|\leq m_h^2$ at $\mu=M_{\Delta}$.}
    \label{Fig:ParamScanNH3}
\end{figure}

\begin{figure}[t]
    \centering
    \subfloat[]
        {\includegraphics[width=0.4\textwidth]{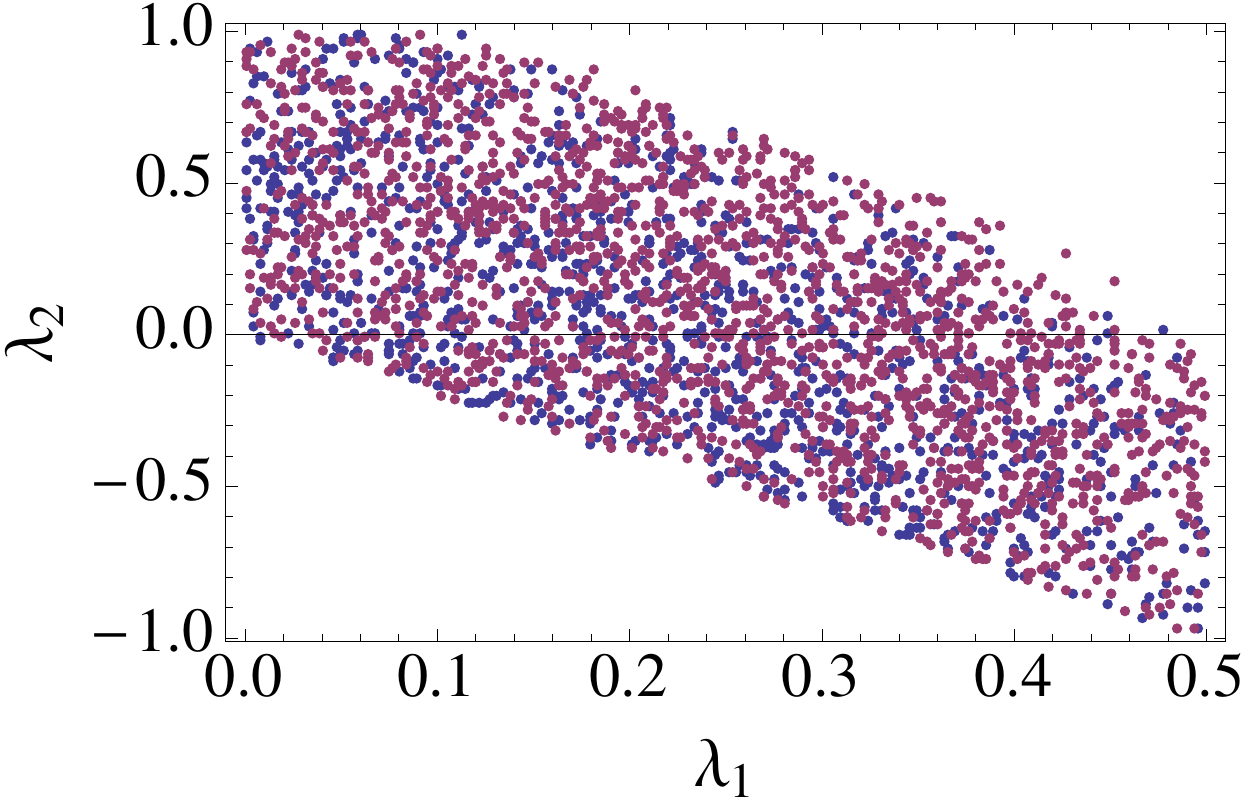}
        \label{Fig:ParamScanNH4a}}
        \qquad
    \subfloat[]
        {\includegraphics[width=0.4\textwidth]{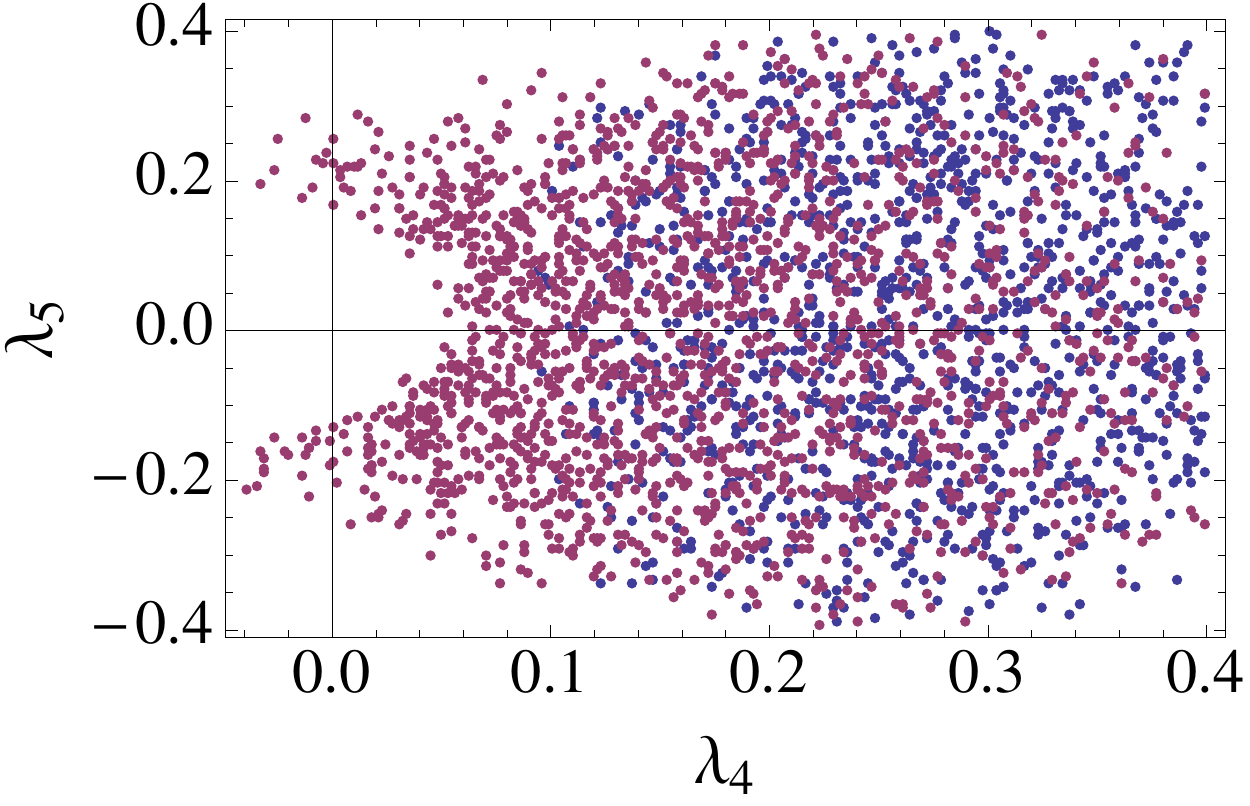}
        \label{Fig:ParamScanNH4b}}
        \\
    \subfloat[]
    {\includegraphics[width=0.42\textwidth]{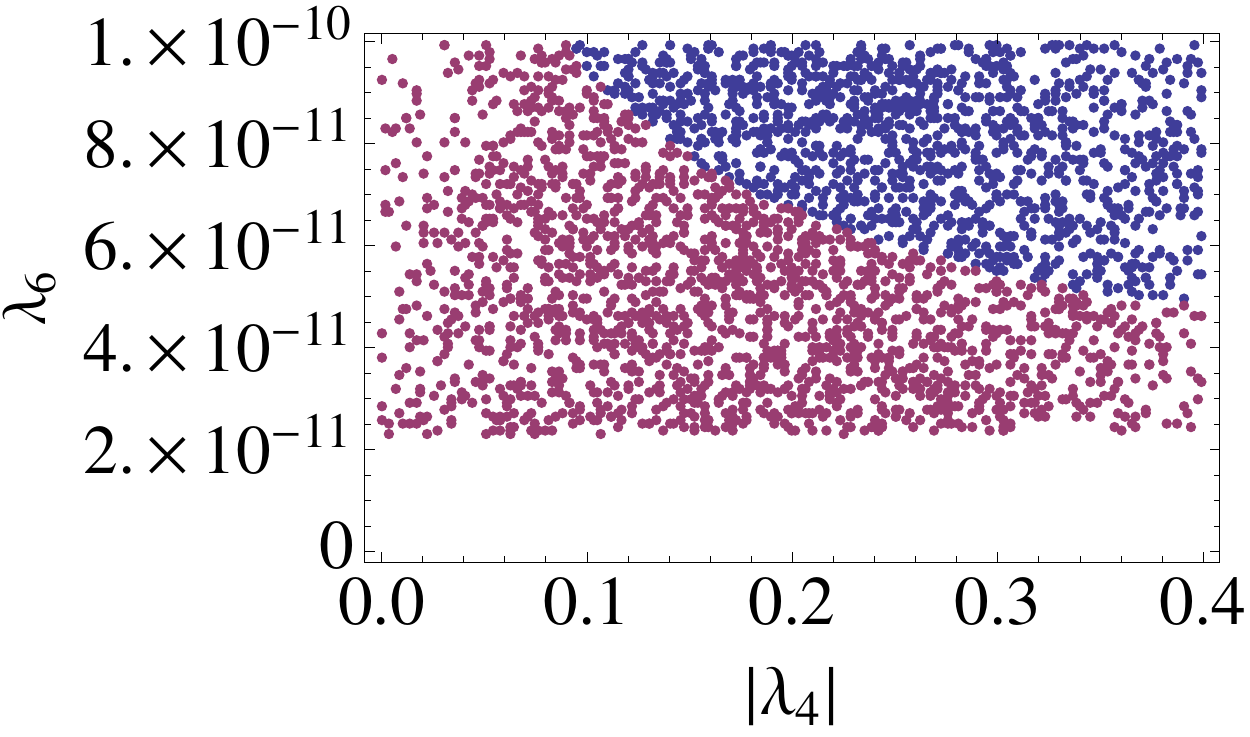}
        \label{Fig:ParamScanNH4c}}
        \qquad
    \subfloat[]
    {\includegraphics[width=0.4\textwidth]{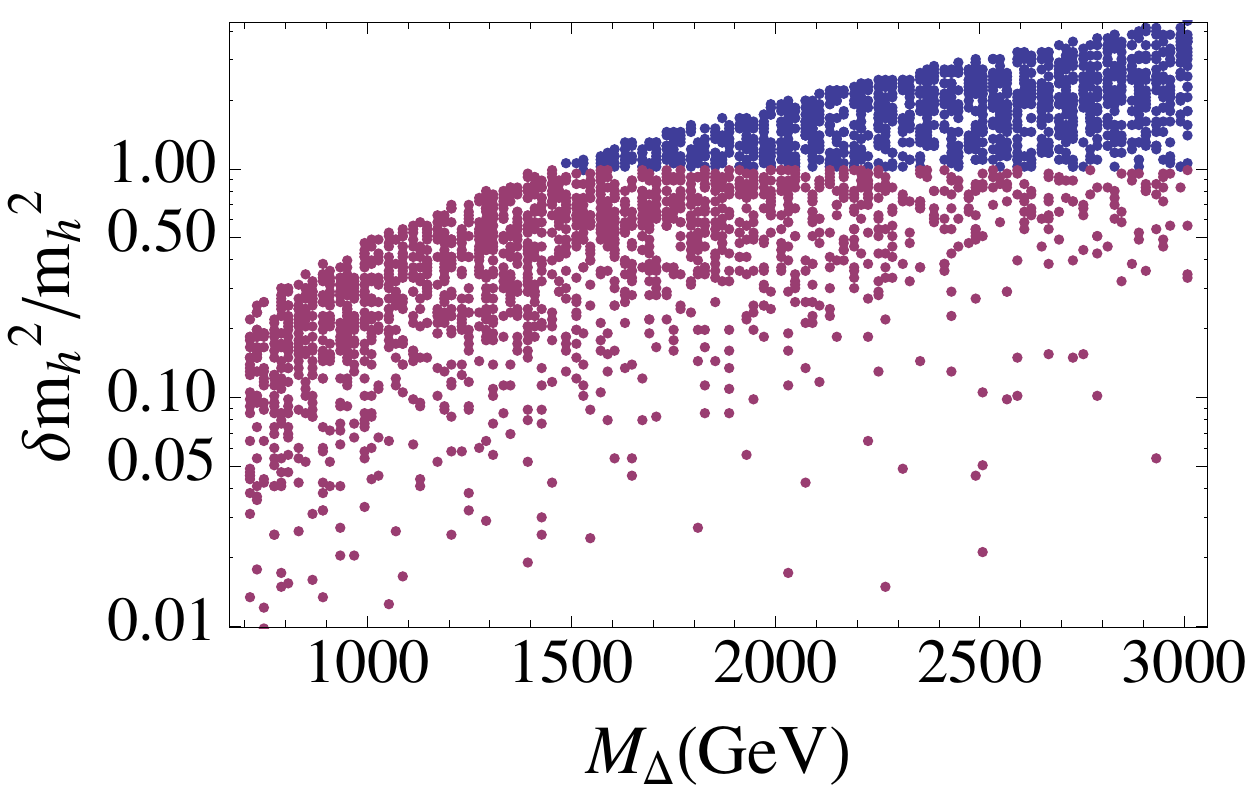}
        \label{Fig:ParamScanNH4d}}
   \caption{Same as Figure~\ref{Fig:ParamScanNH3}, but for $v_{\Delta}=1~\mathrm{eV}$ and $700~\mathrm{GeV}<M_{\Delta}<3~\mathrm{TeV}$.}
    \label{Fig:ParamScanNH4}
\end{figure}

\begin{figure}[t]
    \centering
    \subfloat[]
        {\includegraphics[width=0.4\textwidth]{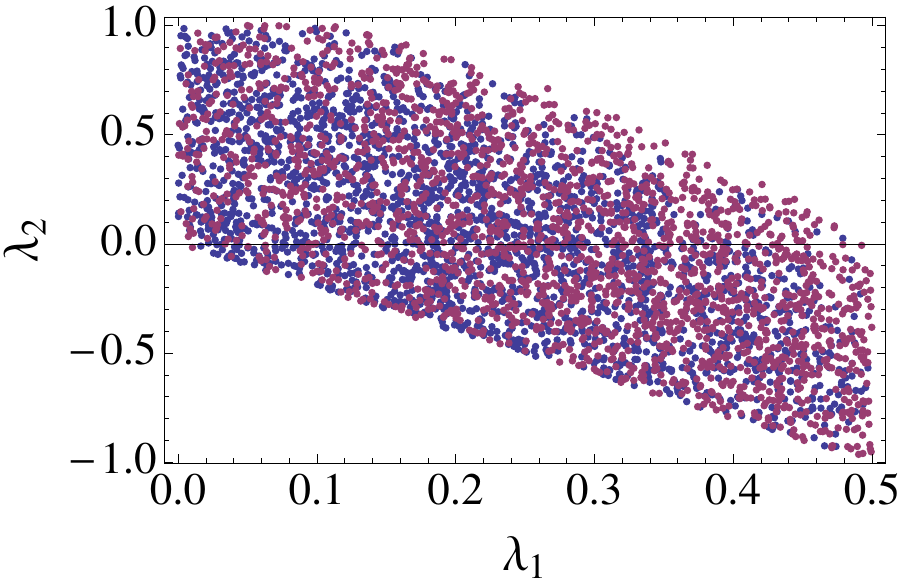}
        \label{Fig:ParamScanNH2a}}
        \qquad
    \subfloat[]
        {\includegraphics[width=0.4\textwidth]{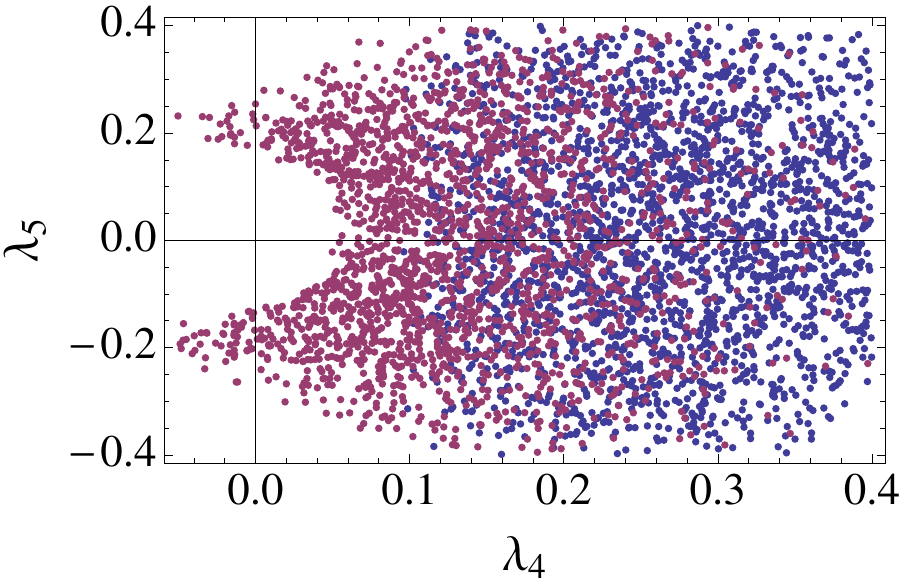}
        \label{Fig:ParamScanNH2b}}
        \\
    \subfloat[]
    {\includegraphics[width=0.42\textwidth]{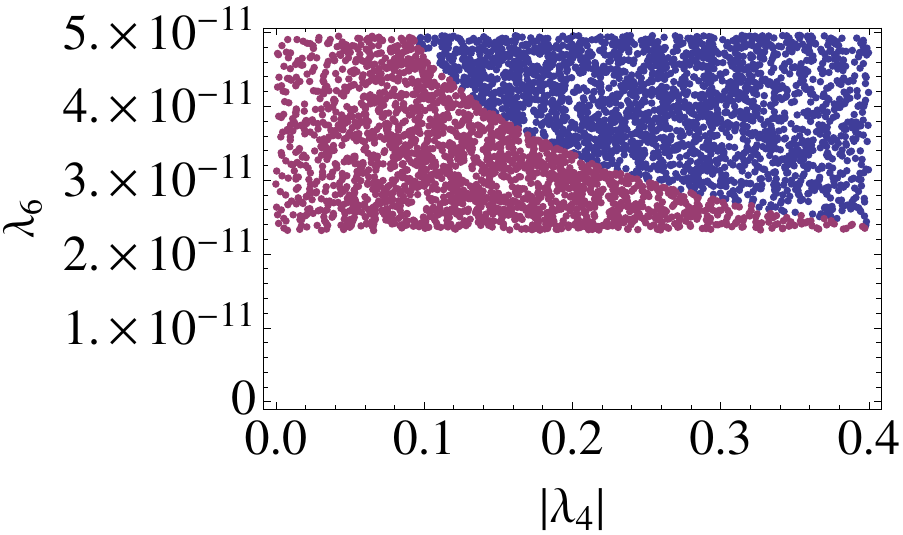}
        \label{Fig:ParamScanNH2c}}
        \qquad
    \subfloat[]
    {\includegraphics[width=0.4\textwidth]{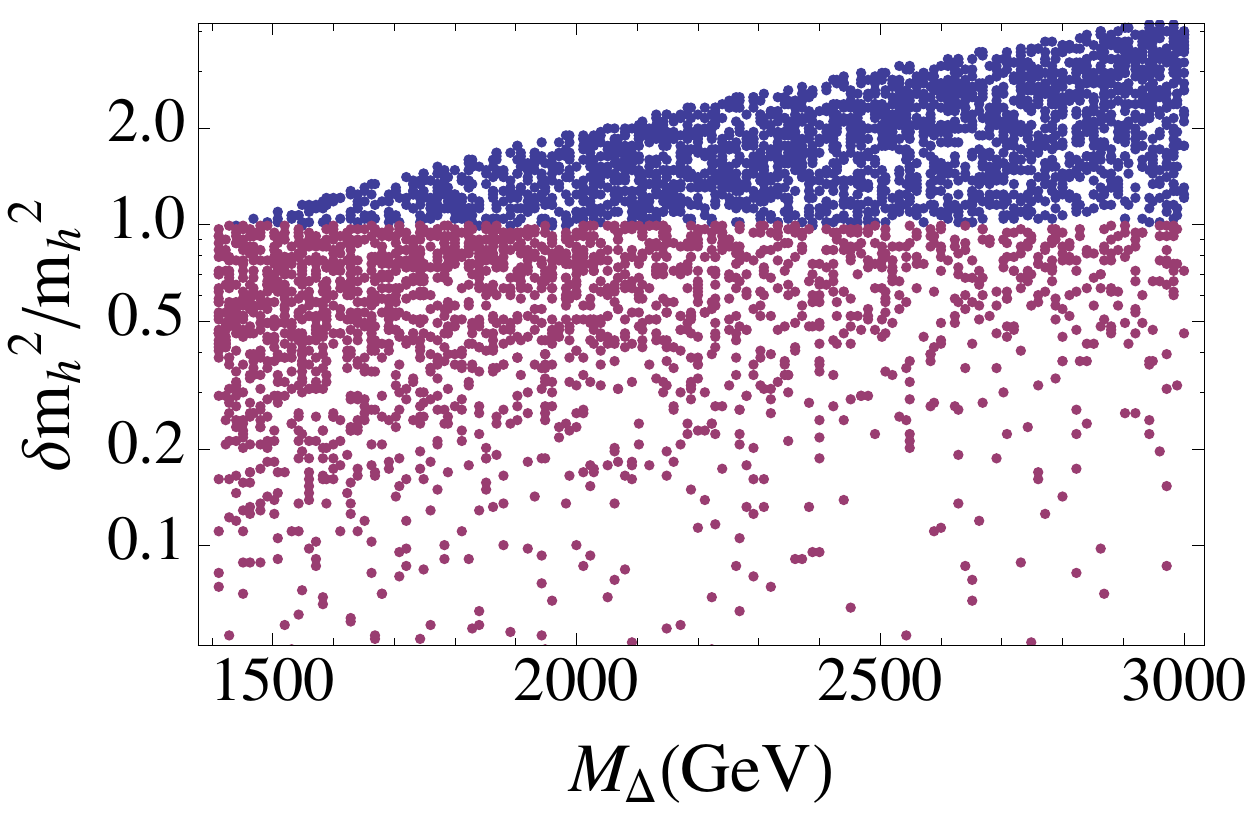}
        \label{Fig:ParamScanNH2d}}
   \caption{Same as Figure~\ref{Fig:ParamScanNH3}, but for $v_{\Delta}=0.5~\mathrm{eV}$ and $1.4~\mathrm{TeV}<M_{\Delta}<3~\mathrm{TeV}$. }
    \label{Fig:ParamScanNH2}
\end{figure}

The parameter $\lambda_6$ is not randomly generated, but is calculated using Eq.~\eqref{Eq:EWSB2} for every set of $(\lambda_4,\lambda_5)$ once $(M_{\Delta},v_{\Delta})$ are fixed. As discussed before, in our scenario $\lambda_6$ is very small in comparison with the other couplings. 

Figures~\ref{Fig:ParamScanNH1}-\ref{Fig:ParamScanNH2} show the allowed parameter space for the type II seesaw model with different choices of $v_\Delta=3.5, 2, 1$ and 0.5 eV, respectively. The blue points in each plot satisfy the perturbativity, vacuum stability and unitarity conditions up to the Planck scale and the pink points also satisfy the naturalness condition $|\delta m_h^2|\lesssim m_h^2$ at $\mu=M_{\Delta}$. 
The experimental bound from direct searches sets $M_{\Delta}\gtrsim 400~\mathrm{GeV}$, while LFV experiments\footnote{Recall that we will only consider the bounds from $\mu \rightarrow e\gamma$, since they are independent of the Majorana phases and the absolute neutrino mass scale. } impose $M_{\Delta}>200~\mathrm{GeV}$ for $v_{\Delta}=3.5~\mathrm{eV}$ and $M_{\Delta}>1.4~\mathrm{TeV}$ for $v_{\Delta}=0.5~\mathrm{eV}$ [see Table~\ref{Table:vMBounds}]. Accordingly, we have considered the mass range $400~\mathrm{GeV}<M_{\Delta}<1~\mathrm{TeV}$ for $v_{\Delta}=3.5~\mathrm{eV}$ and $1.4~\mathrm{TeV}<M_{\Delta}<3~\mathrm{TeV}$ for $v_{\Delta}=0.5~\mathrm{eV}$ and in between for the intermediate cases. For illustration, we have chosen NH for the neutrino masses, with the lightest neutrino mass and the Majorana phases zero. The values shown correspond to the initial values of the parameters, i.e.\ the values at $\mu=M_{\Delta}$. The allowed parameter space for different configurations, such as $m_{\nu_{\mathrm{min}}}\neq 0$ and/or IH for the neutrino masses looks essentially 
identical. This is because the only effect from the Yukawa coupling size 
is in the RGEs, where they play a very small role.

For the low scale seesaw with $M_{\Delta}\lesssim3~\mathrm{TeV}$ and small VEV, $v_{\Delta}\sim\mathcal{O}(\mathrm{eV})$, the parameter scan shows that the parameter space is roughly restricted to the values
\begin{equation}
\begin{split}
0 < \ \lambda_1  <  0.5 \,, \quad 
-4\pi  <   \lambda_2  <  4\pi \,, \quad 
-0.1  <  \ \lambda_4  <  0.5\,, \quad 
-0.4  <  \ \lambda_5  <  0.4\,,
\end{split}
\label{Eq:LambdaValues}
\end{equation}
independently of the hierarchy of the neutrino masses, the values of $m_{\nu_{\mathrm{min}}}$ and the Majorana phases. Nevertheless, due to the vacuum stability and unitarity conditions, not all values in these ranges are allowed, but present various correlations, as we will discuss in the following.

Figure~\ref{Fig:ParamScanNH1a} shows the allowed parameter space in the $(\lambda_1,\lambda_2)$ plane. The restriction on the lower value of $\lambda_1$ is $\lambda_1>0$, while for $\lambda_2$ it is $\lambda_2\geq-2\lambda_1$, which correspond to the second and third vacuum stability conditions, Eqs.\ (\ref{Eq:StabCond2}), (\ref{Eq:StabCond3}) respectively. The upper bounds of $\lambda_1$ and $\lambda_2$ come from the perturbativity condition, i.e.\ imposing that the couplings should be smaller than $4\pi$ up to the Planck scale.

Figure~\ref{Fig:ParamScanNH1b} shows the allowed parameter space in the $(\lambda_4,\lambda_5)$ plane. Since the values of $\lambda_6$ are too small to influence the running of $\lambda$, the only possibility to prevent $\lambda$ from becoming negative at high energies is to have large enough values of $|\lambda_4|$ and/or $|\lambda_5|$. 
Therefore, the region around $(\lambda_4,\lambda_5)=(0,0)$ is forbidden, since the RGE for $\lambda$ in the vicinity of this region is almost identical to its SM RGE, and hence, we would hit the SM vacuum instability $\lambda<0$ below the Planck scale, violating the first stability condition, Eq.~\eqref{Eq:StabCond1}. The fourth and fifth vacuum stability conditions, Eqs.\ (\ref{Eq:StabCond4}), (\ref{Eq:StabCond4}), set a lower and upper bound on $\lambda_5$: $\lambda_5\geq -\lambda_4-\sqrt{\lambda\lambda_1}$ and $\lambda_5 \leq \lambda_4+\sqrt{\lambda\lambda_1}$, which exclude the region of large $|\lambda_5|$ for small $\lambda_4$. Large values of both $\lambda_4$ and $\lambda_5$ are excluded by imposing perturbativity up to the Planck scale.

Figure~\ref{Fig:ParamScanNH1c} shows the scatter plot in the $(|\lambda_4|, \lambda_6)$ plane. As explained before, in the low scale seesaw and for small triplet VEV, $\lambda_6$ takes very small values and its effect both in the RGE of $\lambda$ and in the Higgs mass correction are negligible compared to the other parameters. Indeed, from this plot we observe that $\lambda_6$ is of the order $\mathcal{O}(10^{-11}-10^{-10})$, as expected from Eq.~\eqref{eq:lam6}, while $\lambda_4$ is of order $\mathcal{O}(0.1)$.


Finally, Figure~\ref{Fig:ParamScanNH1d} shows the correction to the Higgs mass, Eq.\ \eqref{Eq:SWII-HiggsMassCorrection}, for different values of $M_{\Delta}$, obtained for the different values of the allowed parameter space. The correction has been normalized to the Higgs mass squared so that the naturalness condition reads $|\delta m_h^2|/m_h^2 \lesssim 1$. As we can see, for $v_{\Delta}=3.5 ~\mathrm{eV}$ and $400~\mathrm{GeV}<M_{\Delta}<1~\mathrm{TeV}$ all the parameter space which is allowed by the perturbativity, vacuum stability and unitarity conditions up to the Planck scale satisfies always the naturalness condition. Once $M_\Delta$ crosses about 1.5 TeV, our naturalness criterion is violated, see Figures~\ref{Fig:ParamScanNH3}d-\ref{Fig:ParamScanNH2}d.

Figures~\ref{Fig:ParamScanNH3}-\ref{Fig:ParamScanNH2} show the same allowed parameter space planes as Figure~\ref{Fig:ParamScanNH1} but for lower values of $v_{\Delta}=2,1,0.5~\mathrm{eV}$, respectively. Lowering the triplet VEV requires the triplet masses to be larger in order to fulfill the LFV bounds. In particular, for $v_{\Delta}=0.5~\mathrm{eV}$ the triplet mass is required to be larger than $1.4~\mathrm{TeV}$.  
The $(\lambda_1,\lambda_2)$ and $(\lambda_4,\lambda_5)$ parameter space are the same compared to the ones for $v_{\Delta}=3.5~\mathrm{eV}$. Similarly, the values of $\lambda_6$  are of the order $\mathcal{O}(10^{-11})$. The main difference appears in the values of the correction to the Higgs mass. 
Since now we are considering larger values of $M_{\Delta}$, it would be possible that for some values of the parameter space the correction to the Higgs mass squared became larger than the physical Higgs mass squared, violating the naturalness condition $|\delta m_h^2|\leq m_h^2$. This is in fact the case, as can be seen in Figures~\ref{Fig:ParamScanNH3}d-\ref{Fig:ParamScanNH2}d: for large values of $M_{\Delta}$ and $\lambda_4$ the naturalness condition is not fulfilled (blue points). Nevertheless, there still exist a large parameter space for the whole range of $M_{\Delta}$ in which the naturalness condition is still satisfied (pink points). However, the larger the mass of the triplet and the value of $\lambda_4$, the smaller the allowed parameter space. Thus, from the naturalness point of view, small values of $M_{\Delta}$ and $\lambda_4$ are favored.

As for $v_\Delta = 3.5$ eV, the dependence of the allowed parameter 
space on the neutrino mass and mixing parameters is only present in the RGEs of the couplings and thus not significant.

\subsection{\label{sec:pred} LFV Predictions}
We have seen that there exists a relatively large parameter space in which the vacuum stability, unitarity and perturbativity conditions are satisfied for masses below $\sim 3~\mathrm{TeV}$. The naturalness condition is also satisfied by a large subset of this allowed parameter space. For masses above $\sim 3~\mathrm{TeV}$, the allowed values from naturalness become more and more restricted, being almost non-existent for triplet masses above $\sim 4~\mathrm{TeV}$. This suggests that, if the type II seesaw model is realized in nature and if one wants to keep the radiative corrections to the Higgs mass under control, masses below $3~\mathrm{TeV}$ would be favored. Using this as our motivation, in this section we study the implications of a low-scale triplet for LFV experiments. In particular, we discuss here the prospects for the future LFV experiments like MEG II~\cite{MEG} and PRISM/PRIME~\cite{Kuno:2005mm}, which will search for BR($\mu\rightarrow e\gamma$) to the level of $10^{-14}$ and $10^{-16}$, respectively, 
and in case of non-observation, will set 
 the most stringent limits on the triplet mass, as discussed in Sec.\ \ref{sec:LFV}. 

\begin{figure}[t!]
    \centering
    \includegraphics[width=0.46\textwidth]{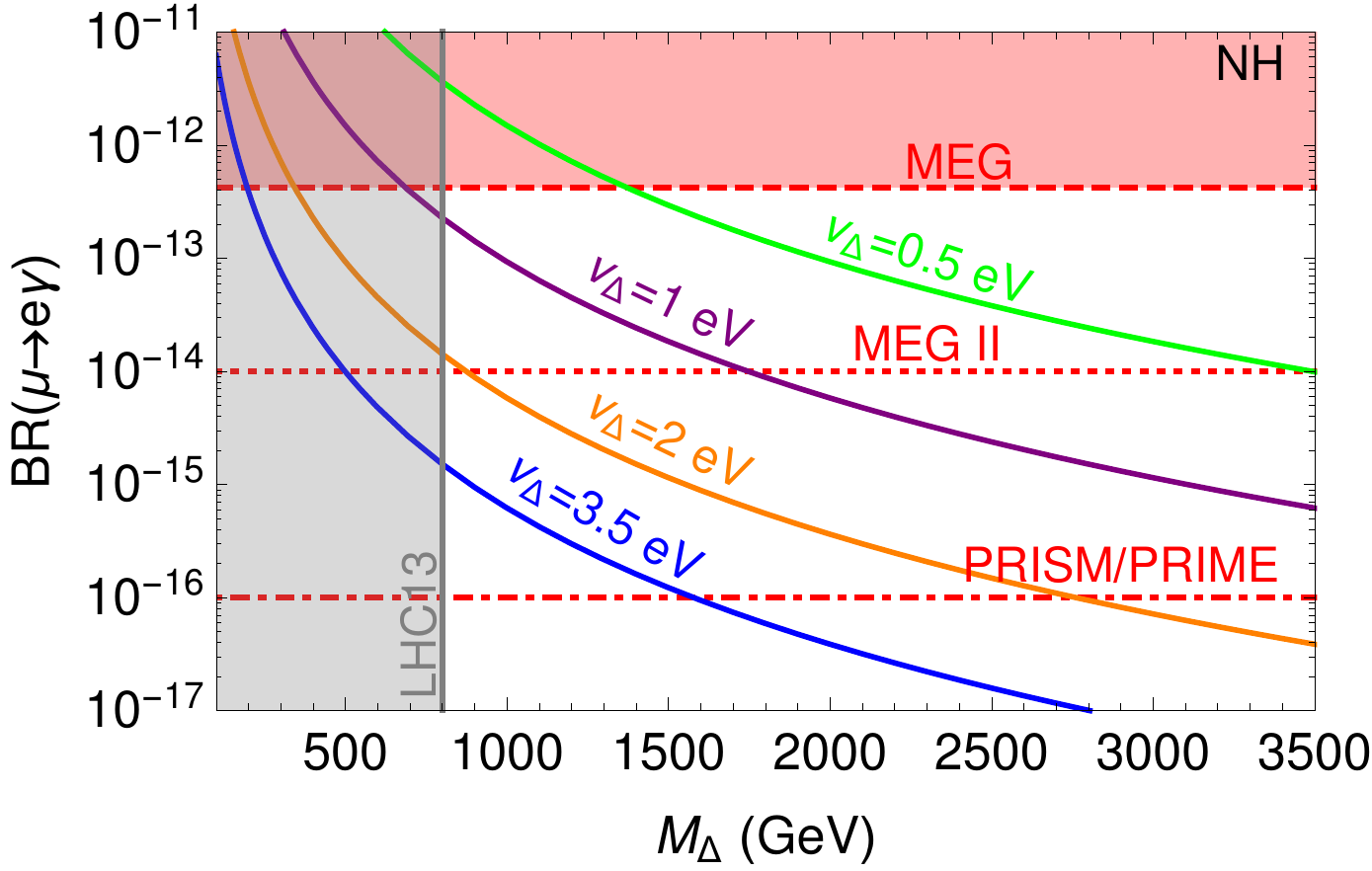}
               \qquad
\includegraphics[width=0.46\textwidth]{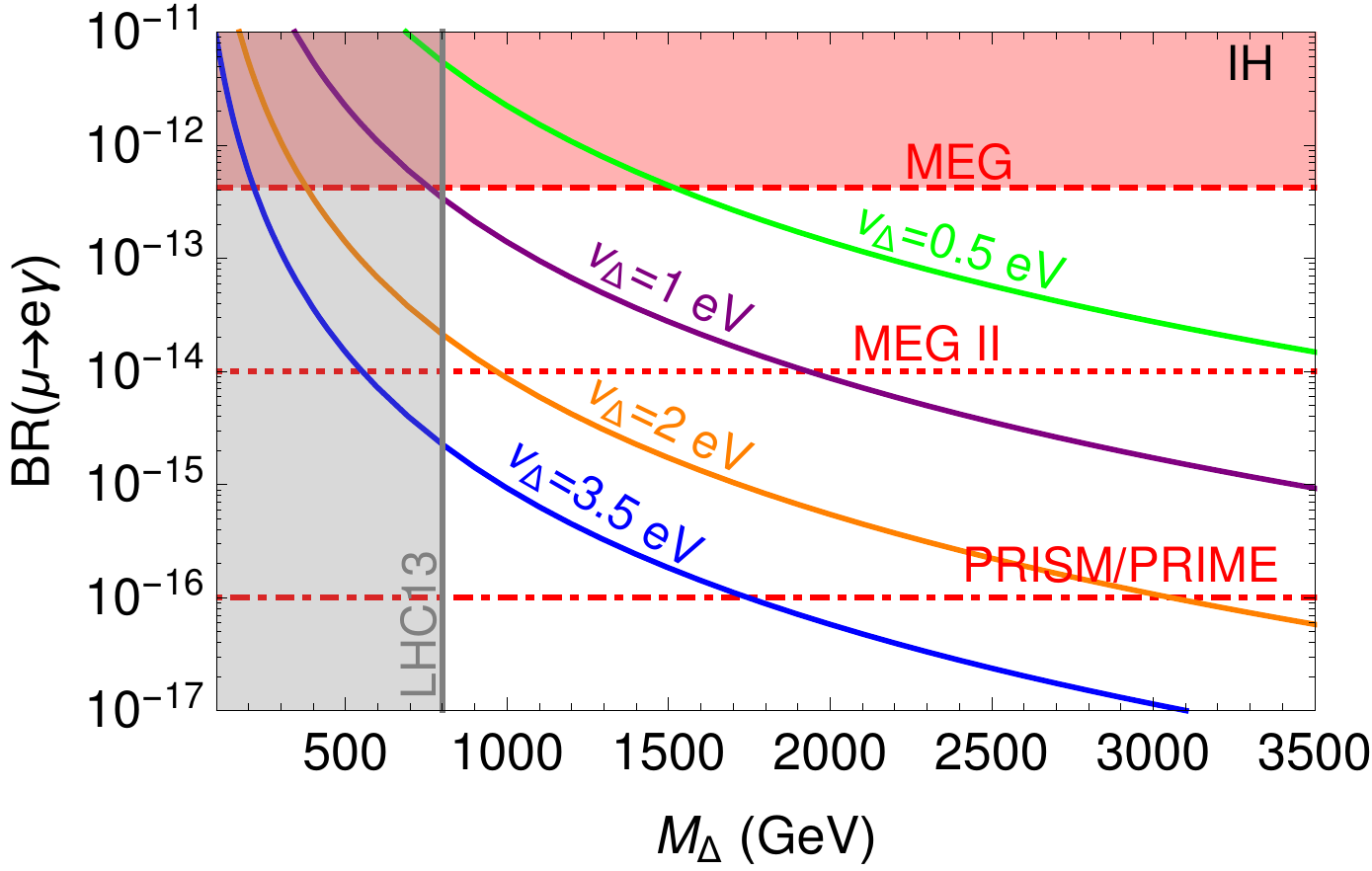}
   \caption{Predicted $\mathrm{BR}(\mu\rightarrow e\gamma)$ in the type II seesaw model for various values of $v_\Delta$. The left (right) panel is for NH (IH). The horizontal shaded (red) region is excluded by the MEG experiment, while the vertical shaded (gray) region is excluded by the LHC. The expected sensitivity of the upgraded MEG II and PRISM/PRIME experiments are also shown.} 
    \label{Fig:BRmuegamma}
\end{figure}

Figure~\ref{Fig:BRmuegamma} shows the dependence of the $\mathrm{BR}(\mu\rightarrow e \gamma)$ predicted by the type II seesaw model [cf.~Eq.~\eqref{Eq:BRmuegamma}] on the triplet mass $M_{\Delta}$ for various values of the triplet VEV $v_\Delta$ and for both NH and IH. Here we have used the best-fit values of the neutrino oscillation parameters from a recent global fit~\cite{Esteban:2016qun}.\footnote{As explained in Section~\ref{sec:LFV}, the branching ratio of this process is independent of the Majorana phases and 
the absolute neutrino mass.}. The horizontal (red) shaded region in each plot denotes the current 90\% CL exclusion from the MEG experiment~\cite{MEG}, whereas the dotted and dot-dashed lines show the expected sensitivity of the upgraded MEG II~\cite{Baldini:2013ke} and PRISM/PRIME~\cite{Kuno:2005mm}, respectively. As can be seen from these figures, for $v_{\Delta}=3.5~\mathrm{eV}$ ($v_{\Delta}=0.5~\mathrm{eV}$) the current limit set by the MEG experiment is $M_{\Delta}\geq 200~\mathrm{GeV}$ ($M_{\Delta} \geq 1.4~\mathrm{TeV}$), while MEG II will be able to probe up to $\sim 500~\mathrm{GeV}$ ($\sim 3.5~\mathrm{TeV}$). The limits are slightly more stringent for IH, as compared to the NH case. The $\sqrt s=13$ TeV LHC limit on the doubly-charged scalars dominantly decaying to electron and muon final states is roughly 800 GeV at 95\% CL~\cite{CMS:2017pet}, as shown by the vertical (gray) shaded region in Figure~\ref{Fig:BRmuegamma}. From these considerations, it follows that 
for $v_\Delta = 3.5$ eV, the $\mu\to e\gamma$ process cannot be detected. But smaller values of the triplet VEV imply larger Yukawa couplings and the LFV decay could be seen in these cases.

\vspace{5mm}

\subsection{\label{sec:pred2}Predictions for $h\to \gamma\gamma$ and $Z\gamma$}
\begin{figure}[t!]
    \centering
        {\includegraphics[width=0.38\textwidth]{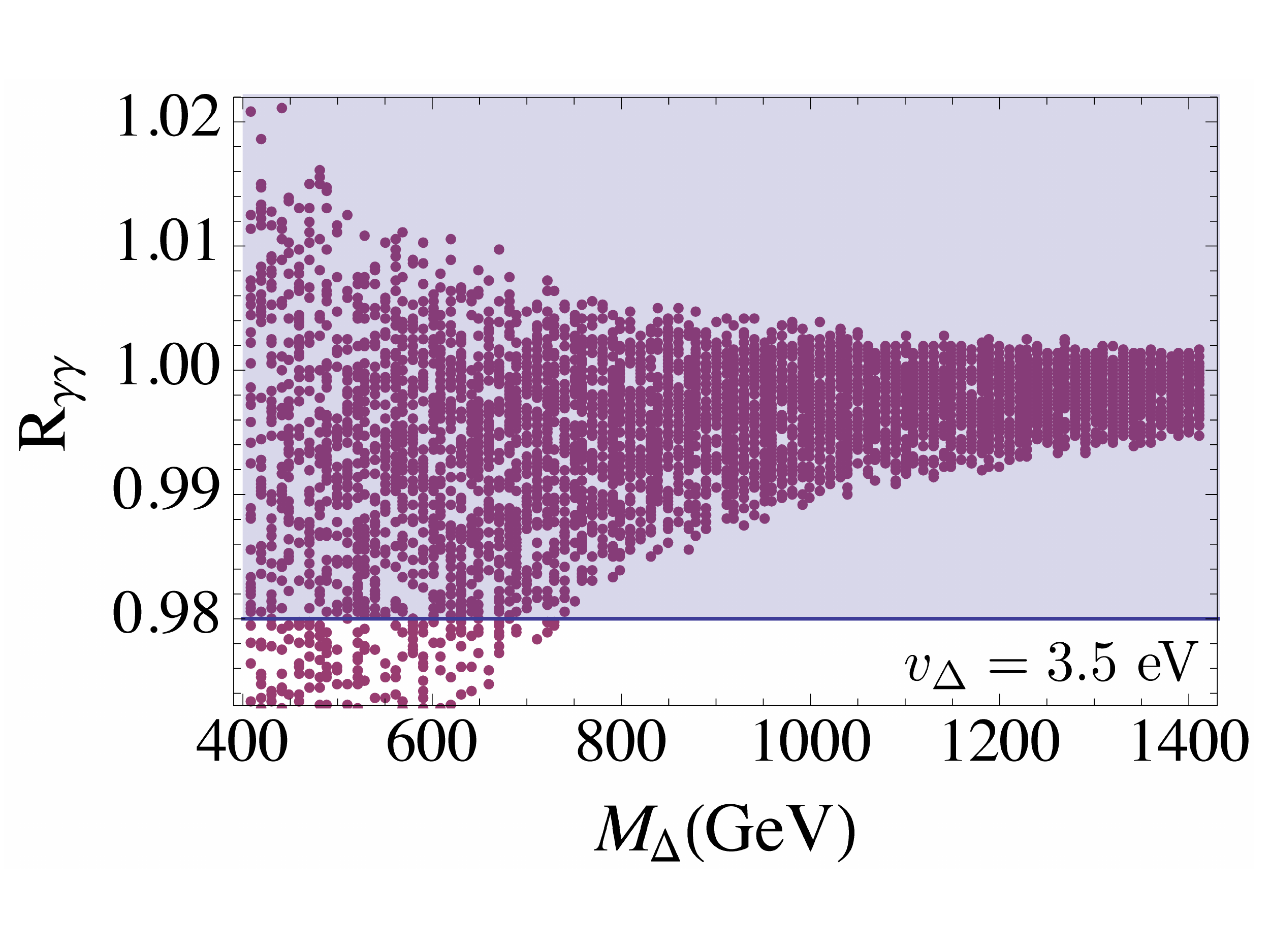}
        \label{Fig:Higgs1}
        }
        \qquad
        {\includegraphics[width=0.38\textwidth]{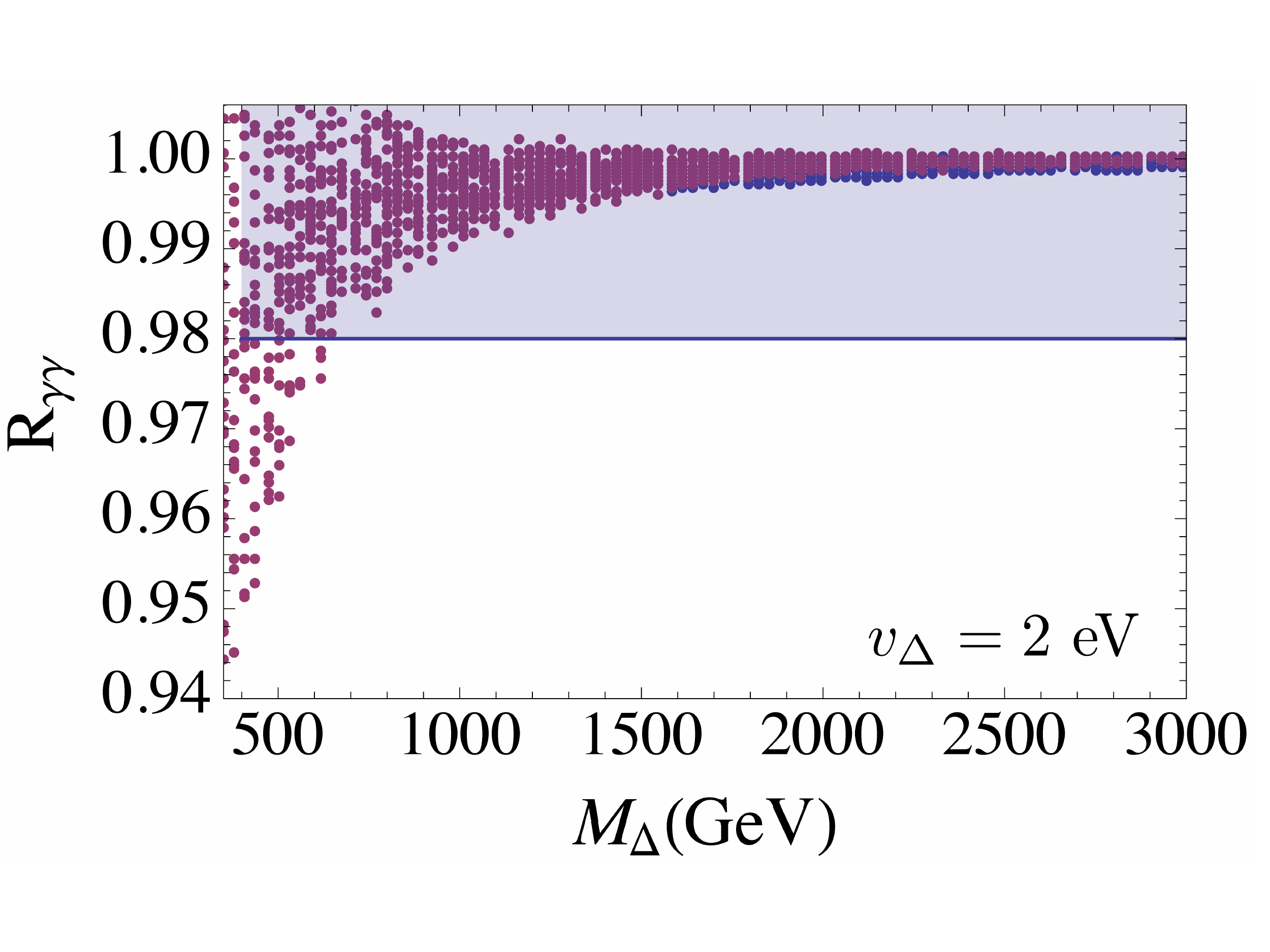}
        \label{Fig:Higgs1}
        } \\ [10pt]
        {\includegraphics[width=0.38\textwidth]{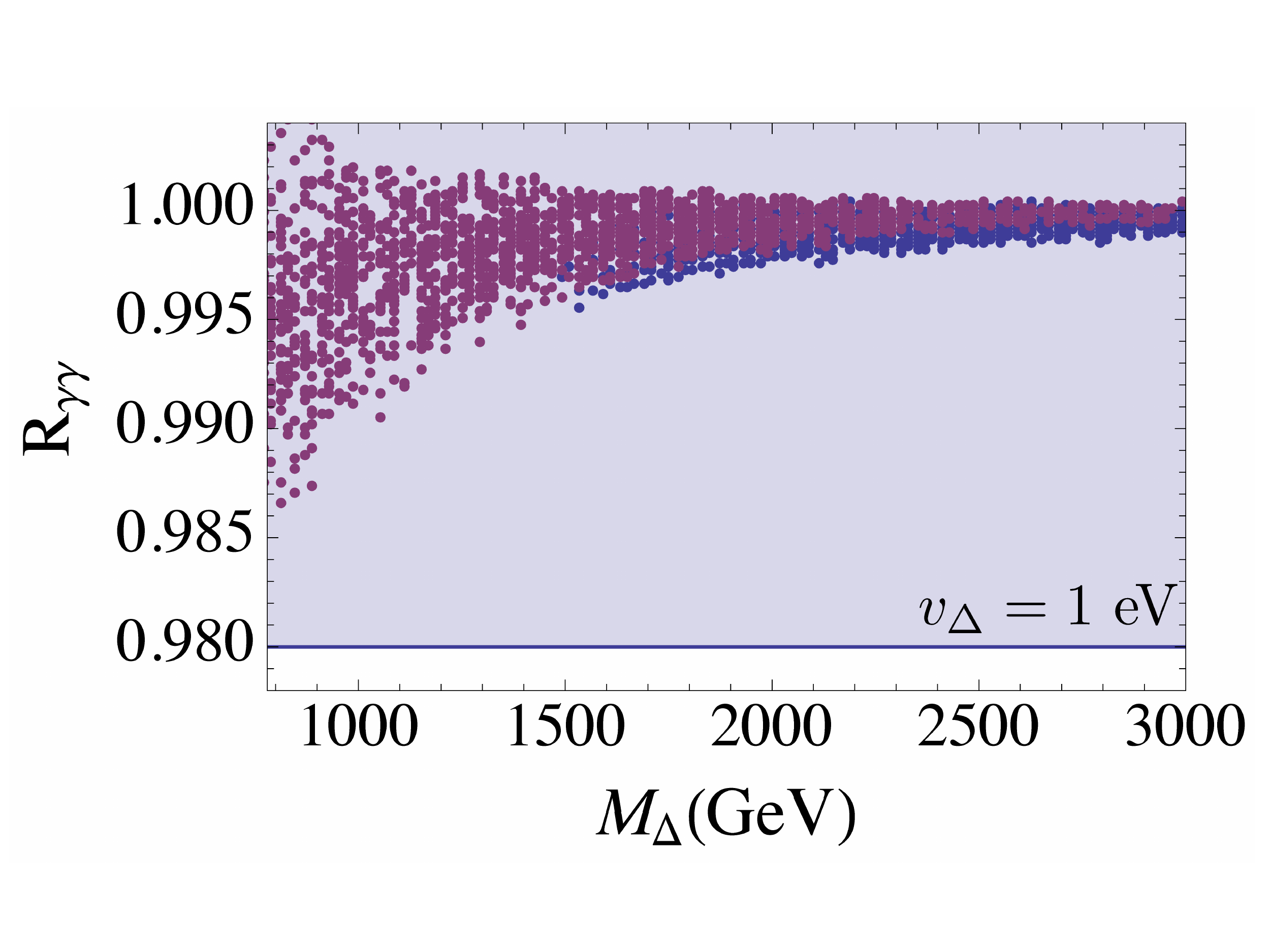}
        \label{Fig:Higgs1}
        }  \qquad 
        {\includegraphics[width=0.38\textwidth]{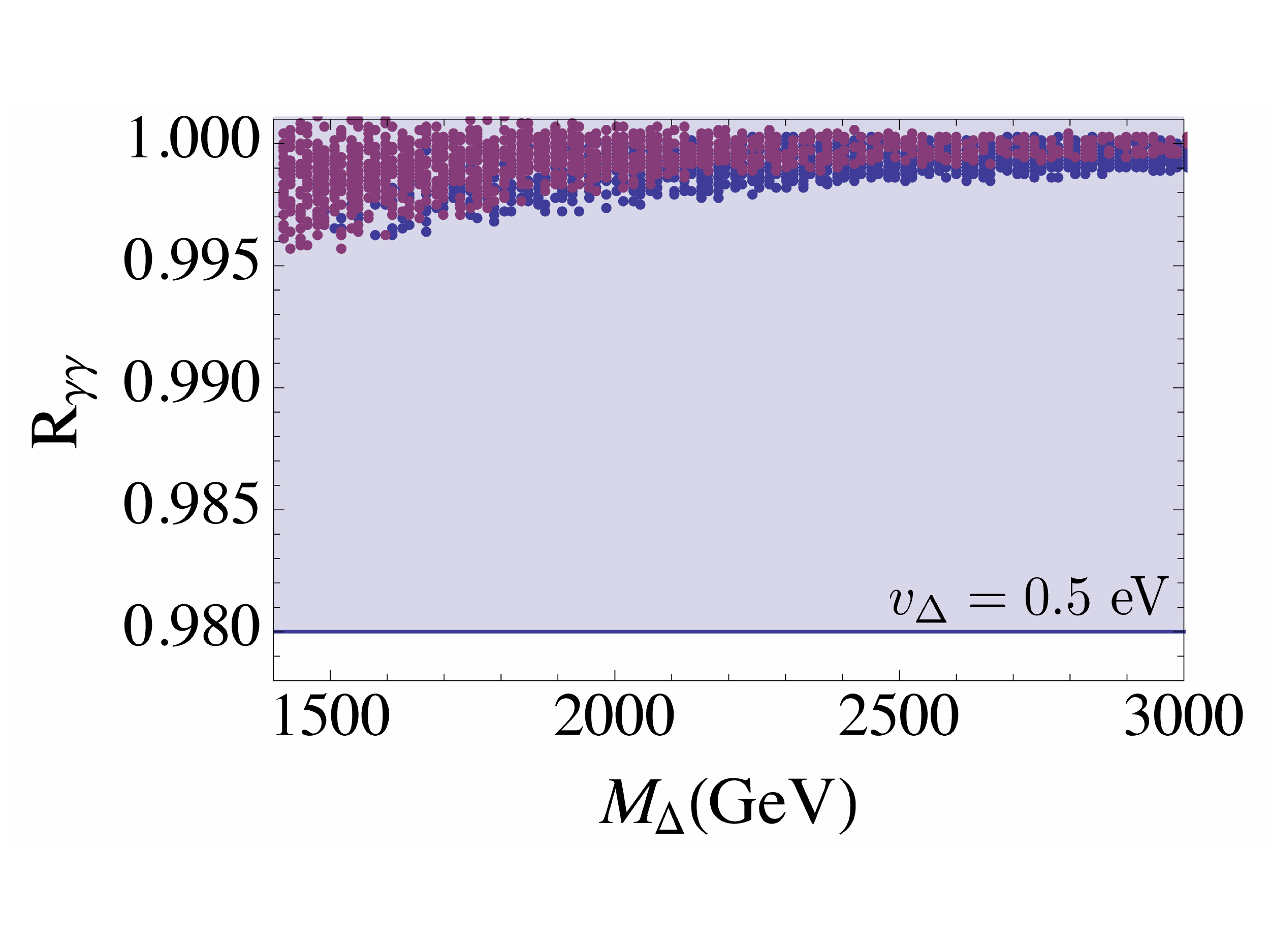}
        \label{Fig:Higgs2}
        }
   \caption{Predicted BR($h \to \gamma\gamma$) with respect to the SM value for $v_\Delta = 3.5, 2, 1, 0.5$ 
 eV. All points satisfy the vacuum stability, unitarity and perturbativity conditions up to the Planck scale. The pink points also satisfy the naturalness condition 
$|\delta m_h^2|\leq m_h^2$ at $\mu=M_{\Delta}$, while the blue ones do not. The blue-shaded 
area corresponds to the experimentally determined range of $R_{\gamma\gamma} = 1.16^{+0.20}_{-0.18}$. \label{Fig:Higgsgg}}
\end{figure}
\begin{figure}[t!]
    \centering
        {\includegraphics[width=0.38\textwidth]{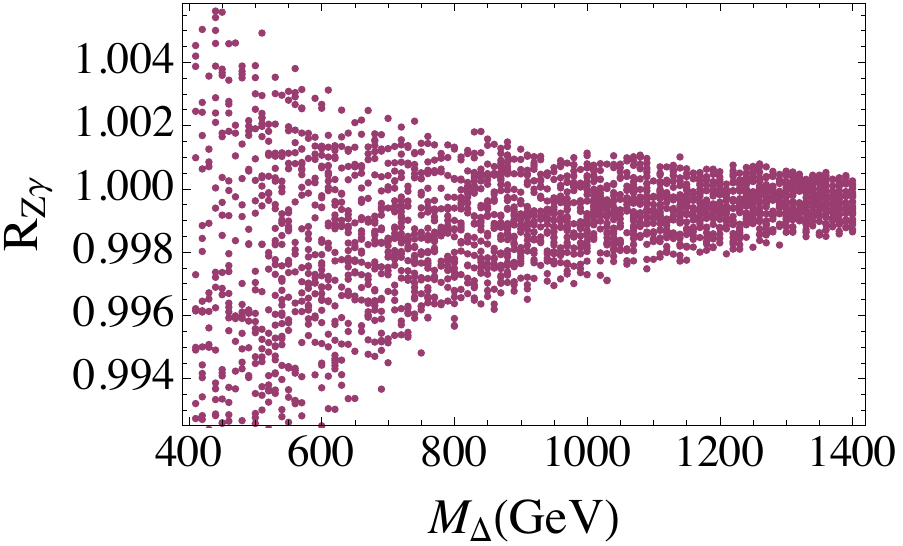}
        \label{Fig:Higgs1}
        }
        \qquad
        {\includegraphics[width=0.38\textwidth]{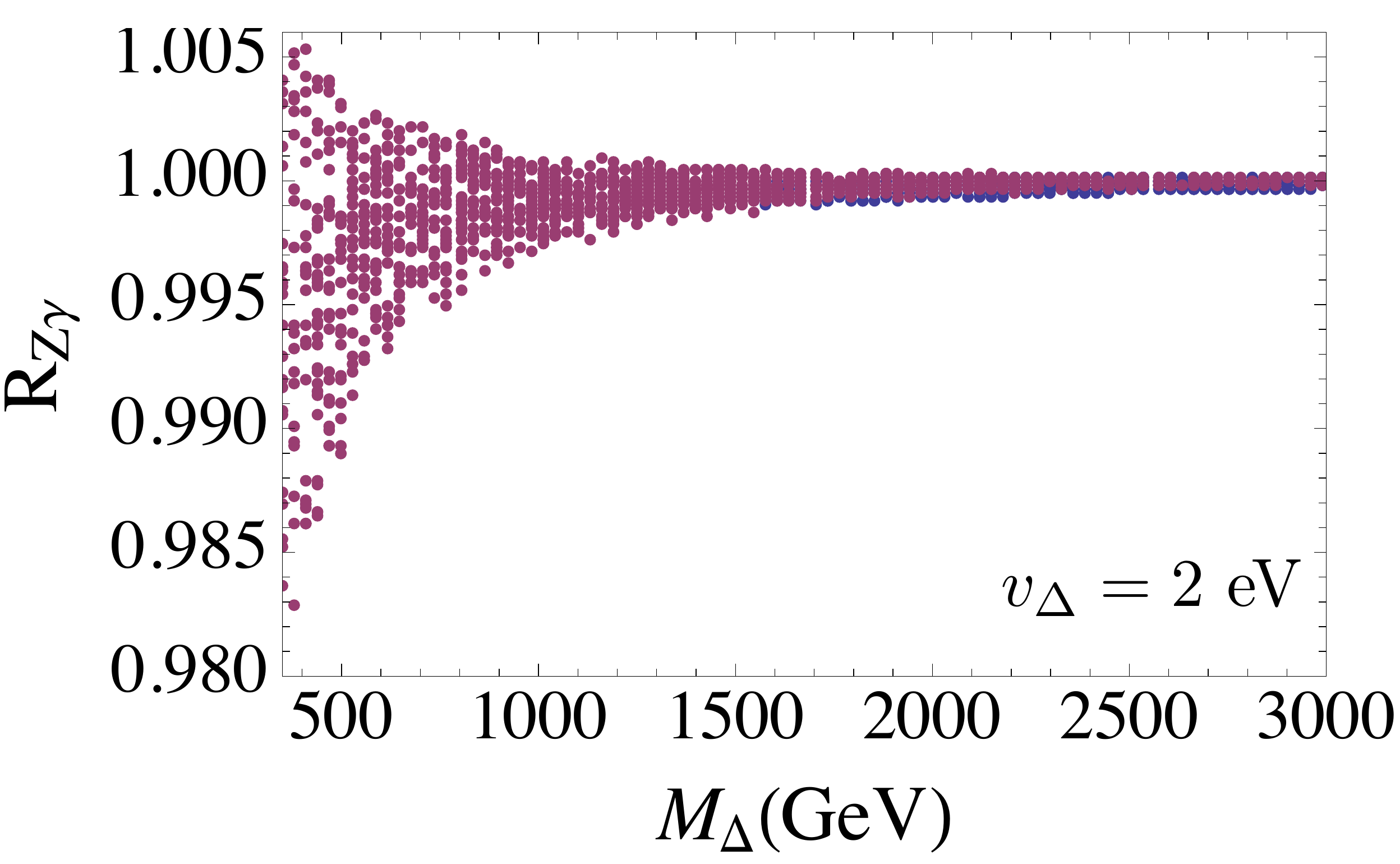}
        \label{Fig:Higgs1}
        } \\ [10pt]
        {\includegraphics[width=0.38\textwidth]{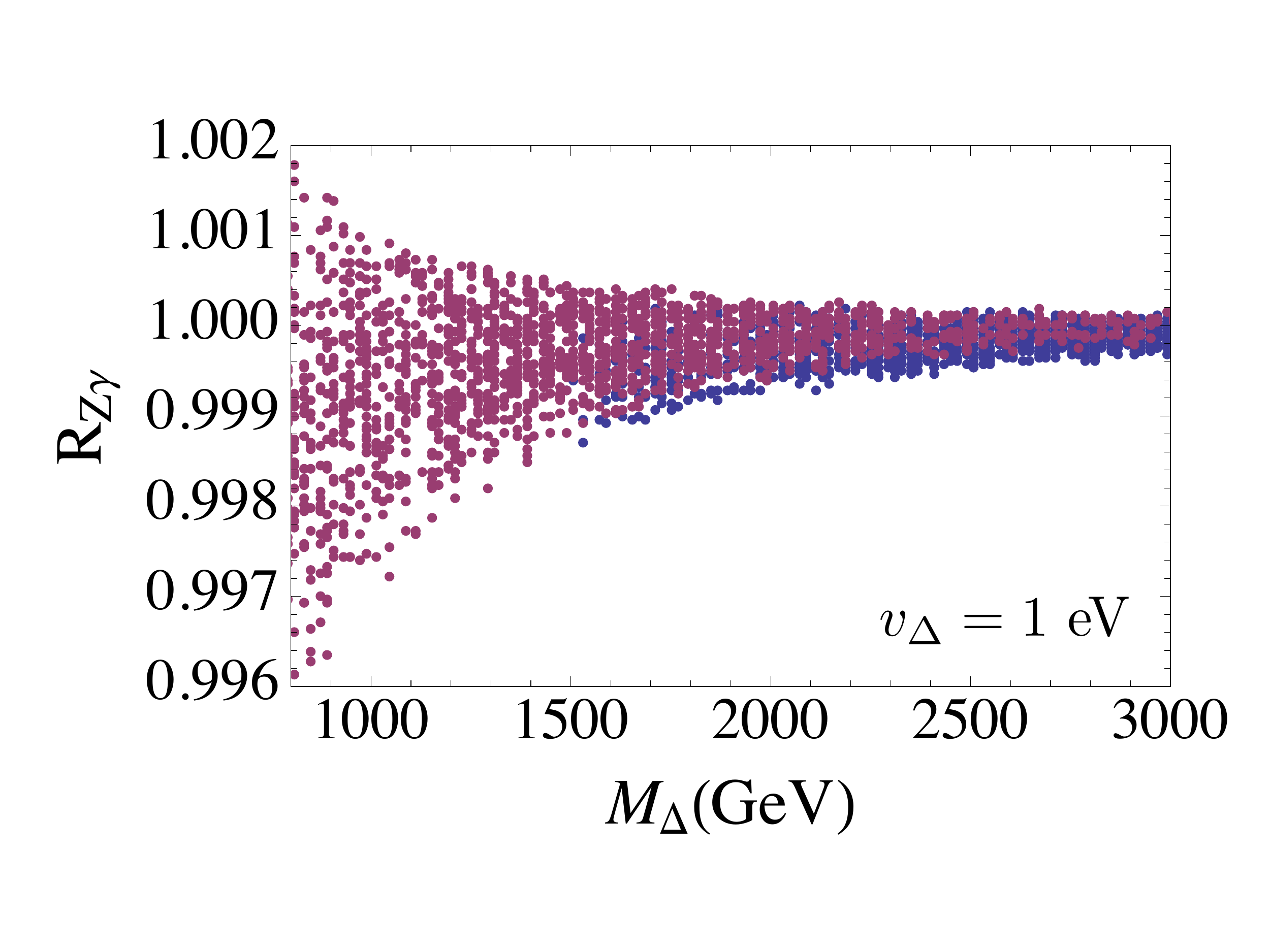}
        \label{Fig:Higgs1}
        }  \qquad 
        {\includegraphics[width=0.38\textwidth]{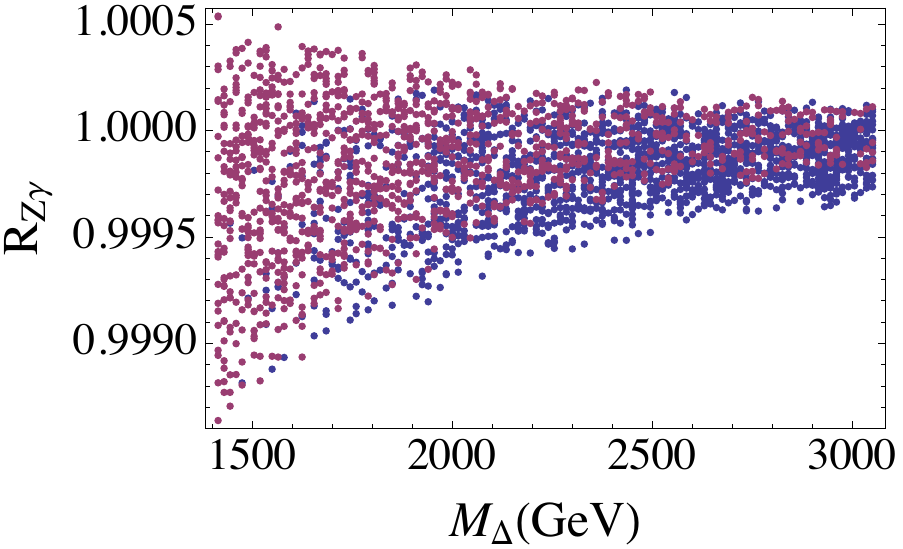}
        \label{Fig:Higgs2}
        }
   \caption{Predicted BR($h \to Z\gamma$)  
 with respect to the SM value for $v_\Delta = 3.5, 2, 1, 0.5$ 
eV.  All points satisfy the vacuum stability, unitarity and perturbativity conditions up to the Planck scale. The pink points also satisfy the naturalness condition 
$|\delta m_h^2|\leq m_h^2$ at $\mu=M_{\Delta}$, while the blue ones do not. The predictions are well within the final HL-LHC sensitivity of $R_{Z \gamma} = 1 \pm 0.30$. \label{Fig:HiggsZg}}
\end{figure}

Another prediction one can make within the low scale type II seesaw is concerning the radiative Higgs decays 
$h \to \gamma \gamma$ and $h \to \gamma Z$. The lengthy expressions for the branching ratios 
can be found in Refs.\ \cite{Dev:2013ff, Chen:2013dh}. They depend on EW parameters, 
triplet masses, $v_\Delta$ and $\lambda_{1,2,4,5}$. Experimentally~\cite{Hgg}, the ratio  
\begin{equation}
R_{\gamma\gamma} = \frac{{\rm BR_{SM}}(h \to \gamma\gamma) }{{\rm BR_{\Delta}}(h \to \gamma\gamma)}
\end{equation}
is constrained to  $R_{\gamma\gamma} = 1.16^{+0.20}_{-0.18}$. The final HL-LHC sensitivity is at around 10\% precision \cite{HgZ}. The analogous ratio $R_{Z \gamma}$ will be measured only 
to within 30\% by the HL-LHC \cite{HgZ}. 

Scatter plots for the ratios $R_{\gamma\gamma}$ and $R_{Z \gamma}$ are shown in Figures~\ref{Fig:Higgsgg} and \ref{Fig:HiggsZg} for our benchmark 
cases of $v_\Delta = 3.5$, 2, 1 and 0.5 eV. As before, all points satisfy the vacuum stability, unitarity and 
perturbativity conditions up to the Planck scale, whereas the pink points additionally satisfy the naturalness condition. We see that the predictions of our testable type 
II seesaw scenario do not allow for any sizable deviations from the SM $h \to \gamma \gamma$ or $h \to \gamma Z$ decay rates. Therefore, this feature could be used to falsify the low-scale type II seesaw model, if a statistically significant deviation in these decay rates is observed in future data.

\section{Conclusions}
\label{sec:conc}
We have studied the type II seesaw model, which accounts for small neutrino masses through the tree level exchange of a heavy scalar triplet. As in any extension of the SM by new heavy particles coupling to the SM Higgs doublet, it 
could lead to a hierarchy problem if the quantum corrections to the Higgs mass were too large. 
A simple naturalness criterion is that the radiative corrections to the Higgs mass should be at most of the 
order of the physical Higgs mass. We have studied the implications of this naturalness criterion on the type II seesaw parameter space. 

We have restricted ourselves to the study to the low-scale scenario, with triplet masses up to the 
$\mathrm{TeV}$ scale, which could be testable at the LHC or future colliders and LFV experiments. We have in addition considered natural 
values of the triplet VEV of the order of $\mathrm{eV}$, which lead to sizable 
Yukawa couplings between the triplet and the SM leptons. In this pragmatic setting, 
our analysis demands that the model parameters obey, 
besides current experimental constraints, vacuum stability, unitarity and perturbativity limits up 
to the Planck scale. With regards to LFV, we have focused mostly on the decay $\mu \to e \gamma$, 
which is guaranteed to happen in the type II seesaw model, unlike other LFV processes like $\mu\to 3e$ which could be suppressed by an appropriate choice of the Majorana phases. 

We have shown that for triplet VEVs larger than 2 eV there is no constraint 
from naturalness if the triplet mass is below 1 TeV. Lowering the VEV below 1 eV implies larger triplet masses 
beyond 1 TeV from the constraint of the decay $\mu \to e \gamma$, and part of the parameter space that is testable 
at the LHC and with $\mu \to e \gamma$ becomes ruled out by naturalness considerations. 
Predictions for $\mu \to e \gamma$ and Higgs branching ratios 
$h \to \gamma \gamma$ and $h \to Z \gamma$ have been made in this setup. In particular, triplet VEVs larger than 2 eV imply no 
signals in $\mu \to e \gamma$ searches for the scenario under study, while smaller VEVs can generate observable LFV decay rates in future experiments. This occurs for a triplet mass regime that is accessible by future colliders.  

Our scenario is thus testable and provides a straightforward example on low scale neutrino mass generation with various implications in the Higgs sector and beyond.

\begin{acknowledgments}
We thank Carlos Yaguna for many useful discussions. 
WR is supported by the DFG with grant 
RO 2516/6-1 in the Heisenberg Programme. 
\end{acknowledgments}

\bibliographystyle{apsrev4-1}
\bibliography{mybib}

\end{document}